\documentclass[aps,twocolumn,groupedaddress,showpacs,pra]{revtex4}
\usepackage{amssymb}
\usepackage{amsmath}
\usepackage{graphicx,epsfig}
\usepackage{color}
\usepackage[ps2pdf,colorlinks]{hyperref}
\usepackage{graphicx}
\usepackage{dcolumn}
\usepackage{bm}
\usepackage{times,mathptm}

\input{Qcircuit}

\newcommand{\tr}{\mbox{Tr}}
\input{tcilatex}
\begin{document}

\title{Global operations for protected quantum memories in atomic spin
lattices}
\author{G.~K.~Brennen$^{1}$}
\author{K.~Hammerer$^{3}$}
\author{L.~Jiang$^{2}$}
\author{M. D.~Lukin$^{2}$}
\author{P.~Zoller$^{3}$}
\affiliation{$^1$ Centre for Quantum Information Science and Security, Macquarie
University, 2109, NSW Australia}
\affiliation{$^{2}$Physics Department, Harvard University, Cambridge, MA 02138, USA}
\affiliation{$^{3}$Institute for Theoretical Physics, University of Innsbruck, and
Institute for Quantum Optics and Quantum Information of the Austrian Academy
of Science, 6020 Innsbruck, Austria}

\begin{abstract}
Quantum information processed in strongly correlated states of matter can provide built in hardware protection against errors. We may encode information in highly non local degrees
of freedom, such as using three dimensional spin lattices for subsystem
codes or two dimensional spin lattices for topologically ordered surface
codes and measurement based codes. Recently, in [L. Jiang et al., Nature Physics {\bf 4}, 482 (2008)] the authors showed how to manipulate these global degrees of freedom using optical lattices coupled to a bosonic degree of freedom via a cavity.  We elaborate on these ideas and recapitulate two approaches to implement many body gates necessary for quantum information processing, both relying on controlled interactions of an ancillary cavity mode with the spin system and single ancilla particles. The main focus of the present paper is to analyze the effect of
imperfections such a cavity decay and collective and individual spin decoherence. We present strategies to fight decoherence by monitoring cavity decay and show that high gate fidelities can be achieved in the strong coupling regime of cavity-QED with state of the art parameters.
\end{abstract}
\maketitle

\section{Introduction}

One way to store quantum information is to prepare it in the degenerate
ground subspace of a many body Hamiltonian. The preparation will be robust
if the logical states are stable under environmental noise. This can be
established if there is a gap $\delta E$ between the code space and its
orthocomplement and if the logical operators themselves are sufficiently non
local in nature. In such a case, at sufficiently low temperature $T$, there
will be a suppression of individual errors by a factor $e^{-\delta E/k_BT}$
and an accumulation of errors that leads to an unrecoverable logical error
will be unlikely. Models for quantum memories and computations based on such codes have been put forward in \cite{Kitaev:03,DKL:02,Bacon:06}.   Means to realize the necessary spin lattice Hamiltonians, carrying the protected degenerate ground subspace, exist in systems of interacting atoms or polar molecules trapped in optical lattices \cite{Duan:03, Micheli:06,Brennen:07}. Yet manipulating such codes is non trivial for the very reason that the states are only coupled by global operations. That is, gates on single logical qubits are realized by multi-qubit gates, which are notoriously hard to implement physically.  Other strategies for a digitized simulation of many body Hamiltonians use quantum circuits or entanglement assistance for teleporation of gates (see e.g. Ref. \cite{Bremner:08}).  However, the quantum circuit approach often takes the system outside of the code space mid-circuit and the entanglement assisted gates require $m$ partite GHZ states to simulate a single summand of an $m$ body Hamiltonian.  

\begin{figure}[tbp]
\begin{center}
\includegraphics[width=\columnwidth]{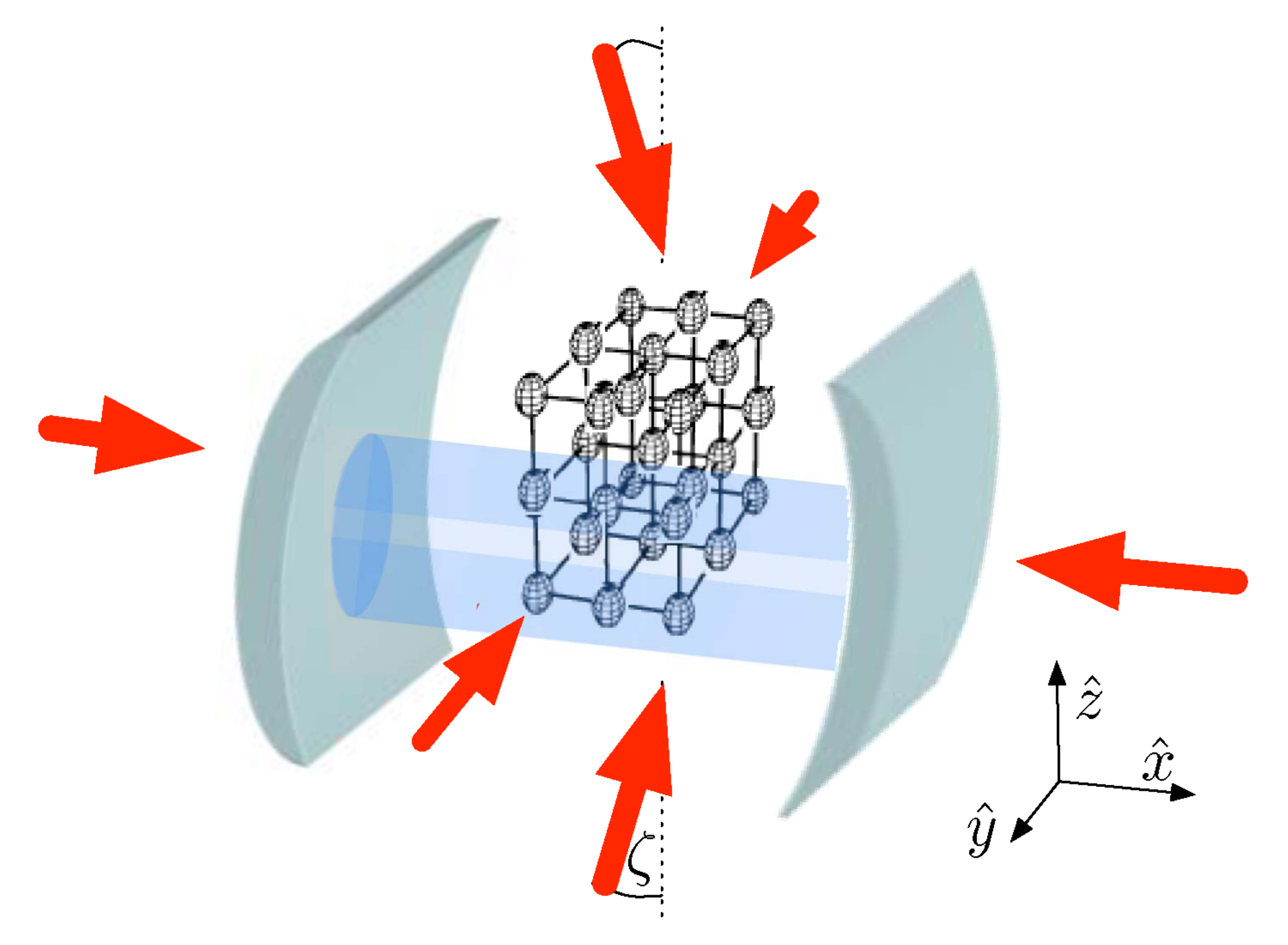}
\end{center}
\caption{Schematic depicting a step in implementing a many body operation on
trapped atoms or molecular spins mediated by a cavity. The lasers for
optical trapping are indicated as is the region of maximum cavity field
strength. Planes of spins can be moved into the strong cavity coupling
region by by stretching the lattice along one dimension (here along $\hat{z}$%
) by e.g. rotating the $\hat{z}$ trapping beams away from the $\hat{z}$ axis by an angle $\zeta$. The
lattice beams are assumed tuned far from cavity resonance so that the
mirrors are transparent at that frequency. One can also use additional
lasers with shaped intensity profiles \protect\cite{Cho:07,Gorshkov:07} to
make a particular region $\mathcal{C}$ of the lattice optically active such
that only spins in that region can interact with the cavity mode.}
\label{fig:1}
\end{figure}

In Ref. \cite{Jiang:07} the authors addressed this challenge suggesting a method to implement the required multi-qubit gates with ultracold atoms or molecules in an optical lattice embedded in a high-quality optical resonator (see Fig. \ref{fig:1}). The key idea is to achieve controlled, collective interactions of selected subsets of spins with the cavity mode interspersed with interactions of a single ancilla particle coupled to the same cavity mode along with measurements of the ancilla qubit. Using these tools two types of gates can be constructed: one is based on the idea to teleport a gate from the ancilla qubit to the encoded qubit and involves single photon excitation of the cavity mode. The other is a geometric phase gate and requires coherent excitations of the cavity mode along with conditional phase rotation controlled by the ancilla qubit. 

While all interactions of atomic (or molecular) spins and the cavity field are obtained in the dispersive limit, cavity decay and spontaneous emission will necessarily affect the gate fidelities. In the present paper we complement our recent proposal \cite{Jiang:07} and provide a detailed analysis of the fidelity of gate operations in the presence of decoherence. We identify the requirements on physical parameters of the system and show that high gate fidelities can be achieved in the strong coupling regime of cavity-QED with state of the art parameters.  Significant experimental progress has been made toward coherent control in this regime \cite{Gupta:07,Colombe:07,Brennecke:07}.  We also identify some adaptive protocols, based on counting of photons leaking out of the cavity, which can improve the gate fidelity in the presence of cavity decay.

The paper is organized as follows.  In Section \ref{Sec:Codes} (supplemented by Appendix \label{AppB}) we introduce examples of protected quantum memories and describe their possible implementations in optical lattices. In Section \ref{Sec:Implem} we present the two gate operations, the single photon protocol and the geometric phase gate. Section \ref{Sec:CavDec} (supplemented by Appendix \label{AppA}) provides a comprehensive treatment of the impact of cavity decay on both protocols along with a discussion of strategies based on monitoring the cavity decay. Section \ref{Sec:OtherDec} finally treats other decoherence effects, in particular individual and collective spin decay.  We conclude with a summary of the results.

\section{Protected quantum memories}\label{Sec:Codes}

We begin with a brief review of the basic properties of ground state quantum memories in three
models: subsystem codes, surface codes, and spin$-1$ chains used for
measurement based computation.

\subsection{Subsystem codes}

\begin{figure}[tbp]
\begin{center}
\includegraphics[height=2.2554in]{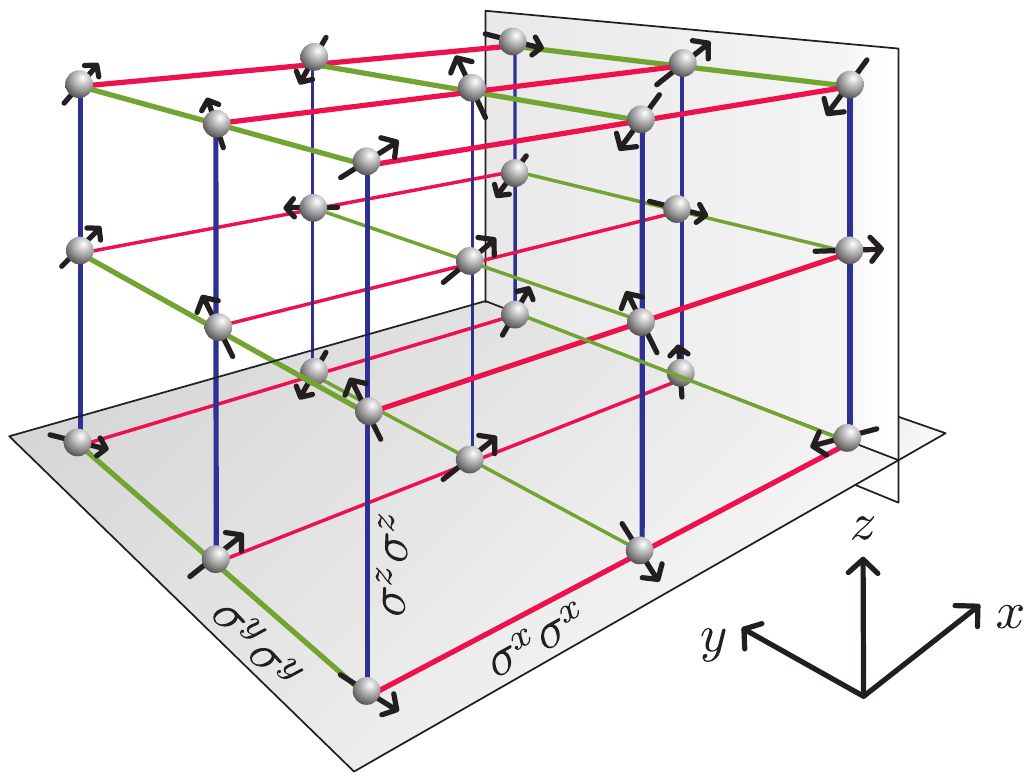}
\end{center}
\caption{A cubic lattice with anisotropic nearest neighbor interaction encoding one logical qubit in $n^3$ physical qubits. Spins interact according to a Hamiltonian $H_{\rm cp}$ with $\sigma^{\alpha}\sigma^{\alpha}$ interactions among each pair of spins in directions $\alpha=x,y,z$. This sytem can be implemented e.g. using microwave induced dipole-dipole interactions between polar molecules as shown in \cite{Micheli:06}. The planar operators $L^X_1$ and $L^Z_1$ which involve a product of $\sigma^x$ and $\sigma^z$ operations along the respective $\hat{y}-\hat{z}$ and $\hat{x}-\hat{y}$ planes act as logical $X$ and $Z$ operations on the code.}
\label{fig:2}
\end{figure}

We focus on a specific subsystem code, the three dimensional spin lattice model studied by Bacon \cite%
{Bacon:06} also known as the $3$D compass model. The system is comprised of qubits residing on the vertices of an
$n\times n\times n$ simple cubic lattice ($n$ odd), as illustrated in Fig. \ref{fig:2}.
The interaction Hamiltonian is:
\begin{equation}
H_{\mathrm{cp}}=-J\big(\sum_{\mathrm{x-links}}\sigma ^{x}\sigma ^{x}+\sum_{%
\mathrm{y-links}}\sigma ^{y}\sigma ^{y}+\sum_{\mathrm{z-links}}\sigma
^{z}\sigma ^{z}\big),  \label{Ham}
\end{equation}%
where $J>0$.  We label the single qubit
operators $\sigma _{i,j,k}^{\alpha }$ for $\sigma ^{\alpha }$ at site $%
x=i,y=j,z=k$.  It is unclear how the gap of this system scales with size 
though numerical evidence for the $2$D compass model suggests that the 
gap scales as $2^{-n}$ \cite{Dorier:05}.  Nevertheless, one could encode 
in a small system to get the benefit of ground state protection.

Stabilizer operators are generated by $2(n-1)$ operators associated with
adjacent planes, including $\sigma ^{x}$ operators in the $\hat{x}-\hat{y}$
plane and $\sigma ^{z}$ operators in the $\hat{y}-\hat{z}$ plane:
\[
V_{i}^{X}=\prod_{j,k=1}^{n}\sigma _{i,j,k}^{x}\sigma _{i+1,j,k}^{x},\quad
V_{k}^{Z}=\prod_{i,j=1}^{n}\sigma _{i,j,k}^{z}\sigma _{i,j,k+1}^{z}.
\]%
We can encode one qubit of information in the ground states of $H_{\mathrm{cp%
}}$, with logical operators $X=\prod_{j,k=1}^{n}\sigma _{1,j,k}^{x}$
(product of operators in the $\hat{y}-\hat{z}$ plane) and $%
Z=\prod_{i,j=1}^{n}\sigma _{i,j,1}^{z}$, product of operators in the $\hat{x}%
-\hat{y}$ plane (see Fig. \ref{fig:2}). Properties of this subsystem code are discussed in
Appendix \ref{AppB}.

\subsection{Surface codes}

Consider the 2D lattice where each edge of the lattice represents the location of a spin-1/2
particle with a coupling graph such that particles on edges which meet a vertex interact via
$H_{v}=\Pi _{j\in \mathrm{{star}(v)}}\sigma _{j}^{x}$ and edges which surround a face interact via 
$H_{f}=\Pi _{j\in
\partial f}\sigma _{j}^{z}$.  The operators $H_v$ and $H_f$ always collide on an even number of
edges and hence commute.  Furthermore, they assume eigenvalues $\pm1$ and information can be encoded in the $+1$ coeigenspace
of these operators.  By choosing the so-called \emph{surface-code
Hamiltonian }\cite{Kitaev:03,DKL:02}:
\begin{equation}
H_{\mathrm{surf}}=-U\sum_{v}H_{v}-J\sum_{f}H_{f},
\end{equation}%
with $U,J>0$ the code space corresponds to the ground subspace with
degeneracy that depends on the topology:  $\mathrm{dim}\mathcal{H}%
_{gr}=2^{2g+h}$ where $g$ is the genus of the surface and $h$ is the number
of holes \cite{DKL:02}. Designing lattices with genus $g>0$, e.g., on the
surface of a torus, would be challenging, but alternatively it would be
possible to create several holes ($h>0$) in a planar lattice with boundary
by for instance deactivating regions of the lattice with focused far detuned
lasers. Furthermore, the planar code with rough boundaries can also
provide a twofold ground state degeneracy of \cite{Bravyi:98}. Since they are insensitive to
local perturbations, the degenerate ground states provide a good quantum
memory.  A caveat is that thermal fragility of the topological order implies a threshold
temperature $T_{\rm thres}$ above which logical information decoheres.  This 
temperature scales as $T_{\rm thresh}\sim \delta_E/\ln n$ \cite{Iblisdir:08} where $n$ is the linear dimension.

\begin{figure}[ptbh]
\begin{center}
\includegraphics[height=2.2554in,width=2.4846in]{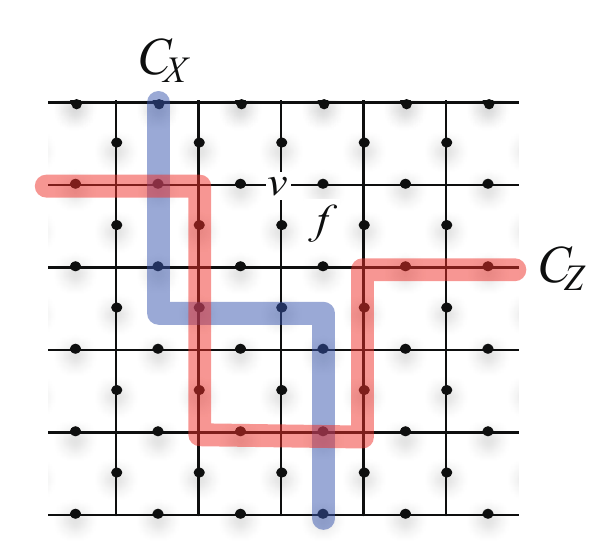}
\end{center}
\caption[fig:6]{A planar code which encodes one logical qubit in the
ground states. There is a spin-1/2 particle (filled dot) for each edge of
the lattice. The interactions of the local Hamiltonian $H_{\mathrm{surf}}$
are along edges that bound a face $f$, and edges that meet at a vertex $v$.
The strings $\mathcal{C}_{X,Z}$ indicate paths of products of $\protect\sigma%
^{x,z}$ operators that are logical operators on the code. }
\label{fig:3}
\end{figure}

The code states are coupled by the logical operators: $Z=\Pi _{j\in \mathcal{C}%
_{Z}}\sigma _{j}^{z}$, and $X=\Pi _{j\in \mathcal{C}_{X}}\sigma _{j}^{x}$
where the configurations $\mathcal{C}_{Z}(\mathcal{C}_{X})$ are strings on
the lattice (dual lattice) as illustrated in Fig.~\ref{fig:3}. %
A more realistic two body Hamiltonian described by 
anisotropic nearest neighbor Ising like interactions on honeycomb coupling
graph was proposed by Kitaev \cite{Kitaev:06} 
which in a certain parameter regime yields an effective Hamiltonian 
unitarily equivalent to $H_{\mathrm{surf}}$. 

\subsection{Ground code computing}
\label{grcode}
Another means of processing information in many body states is via ground
code measurement based quantum computation (GMQC).  Here the information is processed
in degenerate ground states of a gapped Hamiltonian over a two dimensional lattice of spins,
and computation flows by sequential measurement on the constituent spins. It was shown in
Refs. \cite{Gross, BM:08} that it suffices to have a nearest neighbor only interaction between
spin$-1$ particles to realize GMQC with single and two spin measurements.  In the protocol
of \cite{BM:08}, each logical qubit is stored and processsing in the ground states of spin chain with spin$-1$ particles in the bulk and spin$-1/2$ particles on the boundaries interacting via the so-called AKLT Hamiltonian \cite{AKLT:88}
\[
H_{AKLT}=J[\sum_{j=1}^{N-1}P^2_{j,j+1} +P^{3/2}_{0,1}+P^{3/2}_{N,N+1}]
\]
with $J>0$.  Here $P^S_{j,j+1}$ is the projector onto the spin-$S$ irreducible
representation of the total spin for particles $j$ and $j+1$,
i.e. $P^2_{j,j+1} = \frac{1}{2}({\bf S}_{j} \cdot {\bf S}_{j+1} +
\tfrac{1}{3} ({\bf S}_{j} \cdot {\bf S}_{j+1})^2) + \frac{{\bf 1}_9}{3}$ and
$P^{3/2}_{j,j'} = \tfrac{2}{3}({\bf 1}_6 + {\bf s}_{j} \cdot {\bf S}_{j'}),$
where ${\bf S},{\bf s}$ are spin-$1$, $1/2$ representations of
${\mathfrak su}(2)$.
The ground state of $H_{\rm AKLT}$ is non-degenerate but after turning off the interaction
$P^{3/2}_{0,1}$ and measuring the qubit at location $0$, the chain is initialized
into a logical state of a two fold degenerate ground subspace of $H_{AKLT}'=H_{AKLT}-P^{3/2}_{0,1}$.
Computation flows by measuring spins along the the chain and pairs of neighboring spins
in parallel chains with the final output state on the last qubit located at $N+1$.  Such a model is
protected by the hardware from errors but for fault tolerant computation it may be useful to employ multiple chains for each logical qubit using a quantum error correction protocol where the state of the last qubit of one chain is teleported into the ground states of a freshly prepared chain.
The ground
states of $H_{AKLT}'$ are connected by the string operators
$\Sigma^{\mu}
=e^{i\pi \sum_{k=1}^N S^{\mu}_k}\otimes \sigma^{\mu}_{N+1}$.
Hence, given the string initialized in the state induced by measuring the qubit at position
$0_B$ in $\ket{\downarrow_{B_0}}$ we can teleport the logical state of qubit $A$ into
$B$ via the unitary $U=\ket{0}\bra{0}\otimes {\bf 1}+\ket{1}\bra{1}\otimes \Sigma^{x}(j)$.
We describe in Sec. \ref{s1chains} how this could be done by mapping the state of the last qubit
of one chain to a photon, then allowing the photon to interact with the new chain to generate
the many body operator $U$.

\subsection{Universal operations}

In the remainder of this paper we consider how to generate unitary
evolution by the many body operators
\begin{equation}
S_{\mathcal{C}}^{\zeta }=\prod_{j\in
\mathcal{C}}\sigma _{j}^{\zeta },
\end{equation}
 where $\mathcal{C}$ is some set of spins
and $|\mathcal{C}|=m$. We may be interested in performing string operators
on surface codes \cite{Jiang:07} in which case $\mathcal{C}$ is a set which
defines a connected string of spins in a 2D lattice (see Fig.~\ref%
{fig:3}), or perhaps we wish wish to perform encoded operations on
the $[[n^{3},1,n]]$ subsystem code in which case $\mathcal{C}$ is a plane of
spins (see Fig. \ref{fig:2}). One method uses a single photon to generate
the many body interaction and the other uses a geometric phase gate.

Given the ability to generate arbitrary rotations $e^{i\phi
S_{\mathcal{C}}^{X }},e^{i\phi S_{\mathcal{C}}^{Z }}$ and the CNOT gate, and measurements of $Z$, universal
quantum computation can be achieved. For both surface codes and subsystem
codes discussed above, the logical $\mathrm{CNOT}$ operation can be done
transversally between two code blocks. The logical operator $Z$ (or $X$) is
a global operator that couples multiple spins over the encoding block. The
measurement of $Z$ can be achieved by measuring individual physical spins.
However, it is a non-trivial task to perform the logical rotations $e^{i\phi
S^Z_{\mathcal{C}}}$. The center aim of this paper is to provide two approaches to perform
such logical rotations.

\subsection{Physical implementation}

Methods for analogue simulation of the three models above
have been proposed using trapped atoms \cite{Duan:03,Yip:03,Cirac:04} or molecules
\cite{Micheli:06,Brennen:07} in optical lattices.  For trapping of polar
molecules with lattice spacings of $\sim 200$ nm, the interaction strength
using microwave induced dipole-dipole interactions can be as high as
$J\sim 2\pi \times 10$
kHz with decoherence dominated by spontaneous emission of
optical photons at a rate $\gamma \sim 2\pi \times 1$ Hz.
Such small lattice
spacings are possible using trapping lasers tuned to the properly chosen
molecular transitions \cite{Brennen:07}. The implementation $H_{\mathrm{imp}%
} $ will not be exact. There will be spurious longer ranger interactions and
deviations from the required symmetry of the nearest neighbor interaction. Yet such
deviations need not break the code.  For example, the latter two models $H_{\rm surf}$ and
$H_{\rm AKLT}$ have a gap which provides for some resilience to imperfect implementation.
Furthermore, for an implementation of $H_{\rm cp}$ with polar molecules, the microwave fields that induce the interactions are linearly
polarized and the errors are the form of products of pairs of Pauli operators which hence preserve the time reversal symmetry of the model. For this reason, the degeneracy
of the code states is preserved and the gap condition for sufficiently small systems,
 can be maintained
provided the deviations are small. Just how small depends on a detailed
computation of the energy gap to excited states. By optimizing the microwave
beams that induce the dipole-dipole interaction, its found that for a $4$
spin configuration on a $3$D trine, the deviation of the implemented
Hamiltonian $H_{\mathrm{imp}}$ to the target $H_{\mathrm{cp}}$ is: $||H_{%
\mathrm{imp}}-H_{\mathrm{cp}}||<10^{-4}J$, where the norm is defined as the
supremum norm of the traceless part of the operator \cite{Micheli:06}.

\section{Implementations}\label{Sec:Implem}

We focus on physical systems in which an optical lattice is
placed within a high-finesse optical cavity. We assume that we can control
the coupling between the cavity mode and a selected set of spins in the
optical lattice. Such selective manipulation can be achieved using a control
laser beam with appropriately shaped intensity profile \cite{Gorshkov:07,
Cho:07, Jiang:07}. Alternatively, if selected spins form a simple pattern,
such as occupying all the sites in a straight line or a plane, we can
stretch the lattice \cite{Porto:03} and couple the cavity mode with the spins only located
at certain line or plane.

We assume a spin dependent dispersive coupling of our lattice to a cavity
mode:
\begin{equation}
V_{Z}=g_{Z}a^{\dagger }a\sum_{j\in \mathcal{C}}\sigma _{j}^{z},
\label{setofops}
\end{equation}%
where $g_{Z}$ is the dispersive coupling strength, and $\mathcal{C}$ is the
set of selected spins that can be associated with string operator for the
toric code or AKLT model, or planar operator for the subsystem code. 
In addition, it will be
convenient to have a spin dependent coupling of a single ancillary spin to
the cavity:
\begin{equation}
V_{A}=g_{A}a^{\dagger }a\ket{1}_{A}\bra{1}_{A},
\end{equation}%
with coupling strength $g_{A}$, which can be activated by bringing an
ancillary particle into contact with the cavity mode, allowing them to
interact from some time (without interacting with the spin degrees of
freedom of the system particles), and then de-activated by removing the
ancillary particle.

\begin{figure}[tbp]
\begin{center}
\includegraphics[width=8.5cm]{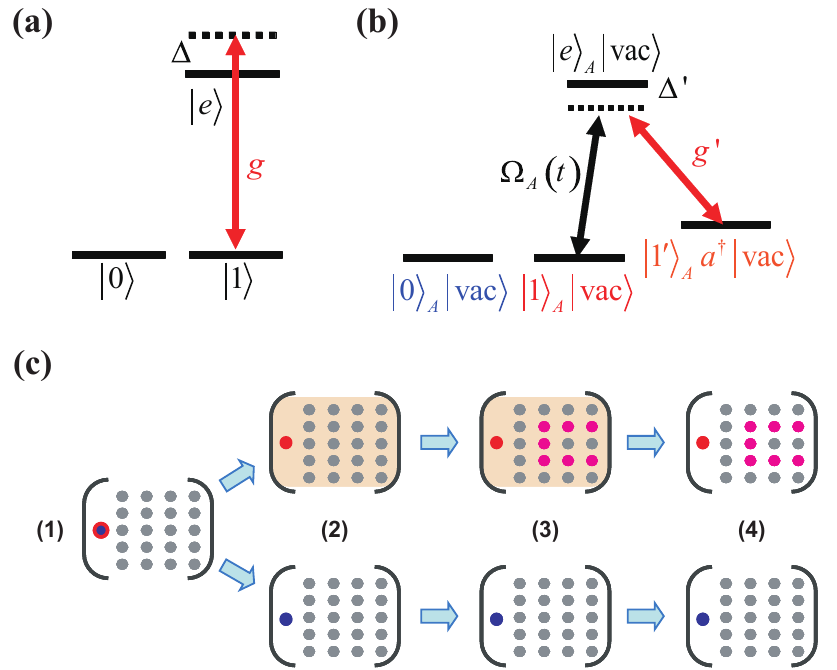}
\end{center}
\caption[fig:4]{Cavity-assisted gate based on
single photon approach. (a) The energy levels of a selected memory spin ($%
\left\vert 0\right\rangle $ and $\left\vert 1\right\rangle $) interacting
dispersively with the cavity mode, which implements the Hamiltonian with
spin dependent dispersive interaction. The coupling coefficient is $\protect%
g_Z=g^{2}/\Delta $, with single-photon Rabi frequency $g$ and detuning $%
\Delta $ from the excited state $\left\vert e\right\rangle $. (b) The energy
levels of the ancilla spin (different from memory spins) and the cavity mode
for the single photon approach. A different control laser with Rabi
frequency $\Omega _{A}\left( t\right) $ connects the states $\left\vert
1\right\rangle _{A}\otimes \left\vert \mathrm{vac}\right\rangle $ and $%
\left\vert 1^{\prime }\right\rangle _{A}\otimes a^{\dag }\left\vert \mathrm{%
vac}\right\rangle $, and enables coherent creation and absorption of a
cavity photon conditioned on the ancilla spin. (c) Cartoon illustration of
the procedure for the implementation of single-photon approach for
controlled many body gate: (1) Initialize the ancilla spin (the left
highlighted spin) in a superposition state $\protect\alpha \left\vert
0\right\rangle _{A}+\protect\beta \left\vert 1\right\rangle _{A}$ (blue for $%
\left\vert 0\right\rangle _{A}$ and red for $\left\vert 1\right\rangle _{A}$%
), with no photon in the cavity and state $\left\vert \protect\psi %
\right\rangle _{S}$ for the topological memory. (2) Coherently create a
cavity photon (orange shade) for ancilla spin state $\left\vert
1\right\rangle _{A}$ (upper branch); no photon is created for ancilla spin
state $\left\vert 0\right\rangle _{A}$ (lower branch). (3) Switch on the
interaction between the cavity photon and the selected spins. If there is a
cavity photon (orange shade), a non-trivial evolution $S_{\mathcal{C}}^{z}$
(pink dots) is implemented. (4) Turn off the interaction and coherently
absorb the cavity photon into the ancilla spin. Finally the state $\protect%
\alpha \left\vert 0\right\rangle _{A}\otimes \left\vert \protect\psi %
\right\rangle _{S}+\protect\beta \left\vert 1\right\rangle _{A}\otimes S_{%
\mathcal{C}}^{z}\left\vert \protect\psi \right\rangle _{S}$ is prepared. }
\label{fig:4}
\end{figure}

The dispersive coupling of the atoms or molecules to the cavity can be
achieved via a state dependent AC stark shift [Fig. \ref{fig:4}(a)].
When the cavity frequency $\omega_c$ is detuned by $\Delta$ far off resonant
from the excited states, then there will be a dispersive
interaction that is photon number conserving. These excited states could be
electronic excited states in the case of an optical cavity or rotational
excited states in the case of a microwave cavity. If, for example, the
photonic mode is $\sigma^+$ polarized then there will be a differential
shift on the $\ket{0}$ and $\ket{1}$ spin states of the polar molecules due
to the different angular momentum coupling coefficients $c_{0,1}$ for the
ground to excited state transitions \cite{Brennen:07}. Up to a constant this
interaction is equivalent to $V_Z$ where $g_Z\sim (c^2_0-c^2_1)\frac{d^2
2\pi\omega_c}{2V\Delta}$ with $V$ the effective mode volume of the cavity
and $d$ the optical dipole moment of the polar molecule. The ancillary
particle, with a different state space such that only one state $\ket{1}_A$
interacted with the cavity mode could be brought into and out of the cavity
with optical tweezers to generate evolution by $V_A$. Recently, a dispersive
interaction between atoms in an optical lattice and an optical cavity was
proposed as a way to measure quantum phases \cite{Mekhov:07}. There the
cavity frequency is chosen far off atomic resonance such that the
interaction is of the type in $V_Z$ except that there is no spin
dependence of the atoms so that the atomic number operator rather than $%
\sigma^z$ is measured. When coupling via an optical cavity it will be
important to choose the lattice spacing along the cavity axis commensurate
with the cavity spatial mode spacing in order to ensure equal coupling to
all spins.

It would be advantageous to have a way to turn on and off the coupling
between the spins and the cavity. This is possible using cavity assisted
Raman pulses. Here the idea is to introduce an auxillary classical field $%
\Omega $ detuned by $\Delta $ from a transition $g\rightarrow e$ and have
the cavity field detuned by $\Delta +\delta $ from a different
transition $g^{\prime }\rightarrow e$. For $\Delta \gg \Omega ,\gamma ,g$
where $\gamma $ is the linewidth of the excited state, then there is an AC
stark $|g_{\mathrm{eff}}|^{2}/\delta $ with $g_{\mathrm{eff}}=\Omega
g (1/\Delta +1/(\Delta+\delta))/2$, and the auxiliary field can turn the coupling on and
off.  It is difficult to find such a closed optical transition
in polar molecules owing to the complex state space which tends to couple
ladders of vibrational levels; although nearly closed transitions do exist
(see \cite{DiRosa:04}). It is possible to find microwave cavity assisted
Raman processes for example by choosing the cavity field to be tuned near
the $N=1\rightarrow N=2$ transition and the auxiliary classical field on the
$N=0\rightarrow N=2$. The latter would have to be a strong field to couple
via quadrapole transitions. Depending on the intermolecular spacing, it may
be necessary to use at least two cavity frequences to obtain only single
spin interactions without inducing spurious dipole-dipole interactions (see
methods of \cite{Micheli:06}).

\subsection{Single photon protocol}

\label{singlephot} Many body gates can be generated with a single cavity
excitation coupling to the lattice. The basic idea is to teleport
the quantum gate from the probe qubit to the encoded qubit of the lattice
spins. One choice for the probe qubit can be the photon number states of the
cavity mode, with zero or one excitation. Alternatively, we may introduce an
ancilla spin as our probe qubit, which couples to the lattice spins via the
common cavity mode. In the rest of the discussion, we will use the second
choice, with the advantage that it is more convenient to manipulate the
ancilla spin compared to the photon number states of the cavity mode.

The procedure of teleporting the quantum gate is summarized in Fig.~\ref%
{fig:5}. First, we couple the probe qubit and the lattice spins. Then,
we perform the quantum gate over the probe qubit. Finally, we measure the
probe qubit, which determines the Pauli frame of the encoded qubit of the
lattice spins. After these operations, the many body gate has been
effectively applied to the lattice spins.

Before we give the procedure of coupling the ancilla spin and the lattice
spins, we first describe a technique that allows use to entangle the ancilla
spin and the cavity mode, using the energy levels shown in Fig.~\ref%
{fig:4}(b). Suppose the ancilla spin and cavity mode starts with
state $\left\vert 1\right\rangle _{A}\otimes \left\vert \mathrm{vac}%
\right\rangle $. When we slowly increase the Rabi frequency of the control
laser $\Omega \left( t\right) $, the ancilla spin and cavity mode
adiabatically follow the dark state $\left\vert \Lambda \left( t\right)
\right\rangle =\left( g^{\prime }\left\vert 1\right\rangle _{A}\otimes
\left\vert \mathrm{vac}\right\rangle -\Omega \left( t\right) \left\vert
1^{\prime }\right\rangle _{A}\otimes a^{\dag }\left\vert \mathrm{vac}%
\right\rangle \right) /\sqrt{\left\vert g^{\prime }\right\vert
^{2}+\left\vert \Omega \left( t\right) \right\vert ^{2}}$, where $g^{\prime
} $ is the single-photon Rabi frequency. Since $\left\vert \Lambda \left(
t\right) \right\rangle $ approaches $\left\vert 1^{\prime }\right\rangle
_{A}\otimes a^{\dag }\left\vert \mathrm{vac}\right\rangle $ for $\Omega
\left( t\right) \gg g^{\prime }$, we effectively transfer the state $%
\left\vert 1\right\rangle _{A}\otimes \left\vert \mathrm{vac}\right\rangle $
to $-\left\vert 1^{\prime }\right\rangle _{A}\otimes a^{\dag }\left\vert
\mathrm{vac}\right\rangle $. Meanwhile, nothing happens if the initial state
is $\left\vert 0\right\rangle _{A}\otimes \left\vert \mathrm{vac}%
\right\rangle $. Since adiabatic state transfer is a coherent process, the
relative phase between $\left\vert 0\right\rangle _{A}\otimes \left\vert
\mathrm{vac}\right\rangle $ and $\left\vert \Lambda \left( t\right)
\right\rangle $ is maintained. Thus, we create an entangled state bewteen
the ancilla spin and the cavity mode
\[
\frac{1}{\sqrt{2}}(\left\vert 0\right\rangle _{A}\otimes \left\vert \mathrm{vac}\right\rangle
-\left\vert 1^{\prime }\right\rangle _{A}\otimes a^{\dag }\left\vert \mathrm{%
vac}\right\rangle) .
\]%
Similarly, we can reverse the state transfer (from $-\left\vert 1^{\prime
}\right\rangle _{A}\otimes a^{\dag }\left\vert \mathrm{vac}\right\rangle $
back to $\left\vert 1_{A}\right\rangle \otimes \left\vert \mathrm{vac}%
\right\rangle $) by adiabatically decreasing the Rabi frequency $\Omega
\left( t\right) $.

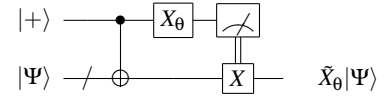
\begin{figure}[ptb]
\begin{center}
\begin{equation}
\Qcircuit@C=1em @R=1em @!R {  &  \lstick{\ket{+}}  &  \qw  &  \ctrl{1}  &
\gate{X_{\theta}}  &  \meter\cwx[1] \\ &  \lstick{\ket{\Psi}}  &  {/}
\qw  &  \targ  &  \qw  &  \gate{X}  &  \qw  &
\rstick{\tilde{X}_{\theta} \ket{\Psi} } \\}\nonumber
\end{equation}%
\end{center}
\caption[fig:5]{Gate teleportation circuit for arbitrary x-rotation $%
\tilde{X}_{\protect\theta}=e^{i\protect\theta S^X_{\mathcal{C}}}$ on the
memory. The circuit represents the following procedure: (1) use
the CNOT gate $\Lambda\left[ \tilde{X} \right] $ to entangle
the probe qubit (upper line) and the memory (lower line with a slash), (2)
projectively measure the probe qubit in a rotated basis, and (3) perform an
encoded Pauli $X$ gate over the memory conditioned on the measurement
outcome. An analogous circuit (using $\Lambda\left[ \tilde{Z} \right] $ and
classically controlled Pauli $Z$ gate) implements $\tilde{Z}_{\protect\theta %
}=e^{i\protect\theta S^Z_{\mathcal{C}}}$, hence by Euler decomposition, any
gate can be formed on the memory using three controlled operations. For the
geometric phase gate scheme, we can actually implement rotations of the
encoded qubit without the probe qubit, e.g., x-rotation of the encoded qubit
can be decomposed as $e^{i\protect\theta\tilde{X}}=D\left( -\protect\beta%
\right) D\left( -\protect\alpha e^{i\frac{\protect\pi}{2}\tilde{X}}\right)
D\left( \protect\beta\right) D\left( \protect\alpha e^{i\frac{\protect\pi}{2}%
\tilde{X}}\right) $, and choosing $|\protect\alpha \protect\beta|=\protect%
\theta$.}
\label{fig:5}
\end{figure}

Using the technique described above, we can implement the controlled
many-body operations [Fig. \ref{fig:4}(c)] as the following:
\begin{itemize}
\item The ancilla spin (probe qubit) starts with state $\alpha
\left\vert 0\right\rangle _{A}+\beta \left\vert 1\right\rangle _{A}$, the
cavity mode has no photon $\left\vert \mathrm{vac}\right\rangle $, and the
topological memory is in state $\left\vert \psi \right\rangle _{S}$.
\item Then we adiabatically turn on the control laser Rabi frequency $\Omega
\left( t\right) $ and coherently create a cavity photon if the ancilla spin
starts in $\left\vert 1\right\rangle _{A}$, so we obtain the state $\alpha
\left\vert 0\right\rangle _{A}\otimes \left\vert \mathrm{vac}\right\rangle
-\beta \left\vert 1^{\prime }\right\rangle _{A}\otimes a^{\dag }\left\vert
\mathrm{vac}\right\rangle $.
\item Next we switch on the interaction between
the cavity photon and selected spins for time $\tau $, during which the
topological memory undergoes an evolution $S_{\mathcal{C}}^{z}$ if there is
one photon in the cavity (i.e. $a^{\dag }\left\vert \mathrm{vac}%
\right\rangle $).
\item Finally, we switch off the interaction and
adiabatically transfer $-\left\vert 1^{\prime }\right\rangle _{A}\otimes
a^{\dag }\left\vert \mathrm{vac}\right\rangle $ back to $\left\vert
1\right\rangle _{A}\otimes \left\vert \mathrm{vac}\right\rangle $.
\end{itemize}
After these four steps, the cavity mode restores the initial state $\left\vert \mathrm{%
vac}\right\rangle $, while the ancilla spin and the topological memory
evolve from $\left( \alpha \left\vert 0\right\rangle _{A}+\beta \left\vert
1\right\rangle _{A}\right) \otimes \left\vert \psi \right\rangle _{S}$ to $%
\alpha \left\vert 0\right\rangle _{A}\otimes \left\vert \psi \right\rangle
_{S}+\beta \left\vert 1\right\rangle _{A}\otimes \left( -i\right) ^{m}S_{%
\mathcal{C}}^{z}\left\vert \psi \right\rangle _{S}$, which is the controlled
operation up to a known phase.

By choosing $S_{\mathcal{C}}^{z}$ as the encoded $Z$ operator, we
implement the controlled-$Z$ gate between the ancilla and the lattice spins,
which can be converted into the CNOT\ gate by conjugating all lattice spins
with the Hadamard gate. Therefore, we can implement all the quantum gates
appeared in the circuit in Fig.~\ref{fig:5} and achieve unitary
evolution of many body operators.


\subsection{Geometric phase gate}

\label{gpimp}

An alternative to using a single Fock excitation of the cavity mode is to use coherent state control to perform a geometric phase gate \cite{Wang:02}.  Since these operations use only linear optical elements,  they may be much easier to realize in experiment. The mechanism makes use of two basic operators, the displacement operator $D(\alpha)=e^{\alpha a^{\dagger}-\alpha^{\ast}a}$ and the rotation operator $R(\theta)=e^{i\theta a^{\dagger}a}$ which satisfy the relations:  
$D(\beta)D(\alpha)=e^{i\Im (\beta \alpha^{\ast})/2}D(\alpha+\beta)$,
and $R(\theta)D(\alpha)R(-\theta)=D(\alpha e^{i\theta})$. Putting these primitives together, one can realize an evolution
\begin{multline}
e^{-iH_{\rm int} t}=D(-\beta B)R(\theta C)D(-\alpha)R(-\theta C)\\
\times D(\beta B)R(\theta C)D(\alpha)R(-\theta C),
\label{seq}
\end{multline}
according to an effective Hamiltonian
\begin{equation}
H_{\rm int}t=|\alpha \beta|\sin(\theta C+\phi),
\end{equation}
where $\phi=\arg(\alpha)-\arg(\beta)$ and $C$ is an arbitrary operator commuting with cavity operators $a$ and $a^\dagger$. Picking in particular $\theta=\pi/2$ and $C=\sum_{j\in\mathcal{C}}\sigma^{\zeta}_j$, which requires driving of the cavity with coherent fields and interactions of the cavity with the spin system according to Eq. \ref{setofops},  the simulated Hamiltonian is
\begin{equation}
H_{\rm int}t=\sin(\phi+m\pi/2)|\alpha \beta|\prod_{j\in\mathcal{C}} \sigma^{\zeta}_j.
\end{equation}
Hence by choosing $C=\sum_{j,k=1}^m\sigma^{\zeta}_{1,j,k}$ and $\phi=0$, for the case $m$ odd, we can simulate evolution generated by $S^{\zeta}_{\mathcal{C}}$.  What is required is a spin dependent coupling between each qubit on the plane and the bosonic channel.

A non destructive measurement of the many body operator $S^z_{\mathcal{C}}$
is possible using the assistance of an ancillary particle $A$ that also
couples the bosonic channel in a spin dependent manner. Controlled
displacements can be implemented by the sequence
\begin{multline}
\ket{0}_A\bra{0}\otimes \mathbf{1}+\ket{1}_A\bra{1}\otimes D(\beta) =
D(\beta/2)R(\pi a^{\dagger}a\ket{1}_A\bra{1}) \\
\times D(-\beta/2)R(-\pi a^{\dagger}a\ket{1}_A\bra{1}).%
\end{multline}
The protocol for measurement of $\langle S^z_{\mathcal{C}}\rangle$ is as
follows

\begin{itemize}
\item Prepare the ancillary particle in the state $\ket{+_y}_A=(\ket{0}_A+i%
\ket{1}_A)/\sqrt{2}$, and the field state in the vacuum.

\item Perform the sequence of steps:
\begin{multline*}
U = D(-\beta/2)e^{-i t_7 V_A}D(\beta/2)e^{-i t_6 V_A}e^{-i t_5 V_Z} \\
 \times D(-\alpha)e^{-i t_4 V_Z}D(\beta/2)e^{-i t_3 V_A}D(-\beta/2) \\
 \times e^{-i t_2 V_A}e^{-i t_1 V_Z}D(\alpha),%
\end{multline*}
which returns the cavity to the vacuum state. Choose parameters satisfying: $
-g_A t_2=g_A t_3=-g_A t_6=g_At_7=\pi$, $-g_Z t_1=g_Z t_4=-g_Z t_5=\pi/2$,
and $|\alpha\beta|=\frac{\pi}{2}\bmod{ 2\pi}$. The displacement operators are
generated by $V_B$ and we pick the relative phase $\phi=\arg{\alpha}-\arg{
\beta}=\pi/2(0)$ for the number of spins $m=|\mathcal{C}|$ even(odd).
Reversed evolution during the interaction time steps $t_1,t_2,t_5,t_6$ can
be achieved by setting $g_{Z,A}\rightarrow -g_{Z,A}$, e.g. by changing the
sign of the detuning.

Assuming the system state was initially $\ket{\psi}_S$ the joint state of
system and ancilla is now
\[
U\ket{\psi}_S\ket{+_y}_A=\frac{\ket{\psi}_S\ket{0}_A+f(m)S^z_{\mathcal{C}}%
\ket{\psi}_S\ket{1}_A}{\sqrt{2}},
\]
where $f(m)=(-1)^{\frac{m}{2}}$ for $m$ even and $f(m)=(-1)^{\frac{m-1}{2}}$
for $m$ odd.

\item Measure the ancilla in the basis $\ket{\pm x}_A=(\ket{0}_A\pm \ket{1}%
_A)/\sqrt{2}$. The probabilities for the measurement outcome $s_x=\pm 1$ are
\[
P(s_x=\pm 1)=\frac{1}{2} (1\pm f(m)\langle S^z_{\mathcal{C}}\rangle).
\]
\end{itemize}

This protocol can be used to measure the product operators $S^x_{\mathcal{%
C^{\prime }}}S^z_{\mathcal{C}}$, where $\mathcal{C},\mathcal{C^{\prime }}$
are possibly overlapping configurations of spins, via the iterated sequence:
$[\prod_{i\in\mathcal{C^{\prime }}}H_i]U[\prod_{i\in\mathcal{C^{\prime }}%
}H_i]U$ with $H_i=e^{i\frac{\pi}{2} (\sigma^x_i+\sigma^z_i)/\sqrt{2}) }$ the
Hadamard gate on qubit $i$. Measurement errors can be ameliorated by
redundifying the state of the ancilla before measurement using many,
possible faulty, $\mathrm{CNOT}$ gates offline between that ancilla and many
others prepared in state $\ket{0}$.

\subsection{Gates on spin$-1$ particles}
\label{s1chains}
We conclude this section with a brief discussion of generating many body
gates for teleportation of quantum information in spin$-1$ chains
as prefaced in Sec. \ref{grcode} above.
This could be done in the context of a lattice embedded in a cavity by mapping the state of the qubit located at position $N+1$ of  logical chain $A$
to the state of a photon and using the same procedure as described above.
The goal is to then generate the unitary operator
$U=\ket{0}\bra{0}\otimes {\bf 1}+\ket{1}\bra{1}\otimes \Sigma^{z}$,
 where $\Sigma^{\mu}
=e^{i\pi \sum_{k=1}^N S^{\mu}_k}\otimes \sigma^{\mu}_{N+1}$,
on a new chain $B$ initialized in state $\ket{\downarrow_{B_0}}$.
This is generated by an identical procedure to that for controlled string operators on
qubits but instead of Eq. \ref{setofops}, we use
\begin{equation}
V_{X}=g_{Z}a^{\dagger }a(\sum_{j=1}^N \ket{S^x=0}_j\bra{S^x=0}+\sigma^x_{N+1}),
\end{equation}%
where $g_{Z}$ is the dispersive coupling strength.  This kind
of state dependent interaction between a photon and spin$-1$ particles
can be realized using e.g. polarization to differentially couple the
internal states of the particles.

\section{Gate fidelity with cavity decay}\label{Sec:CavDec}

Cavity field decay at a rate $\kappa$ acts as a source of 
error for the many body
interactions which it mediates. For the protocols above where the system of spins interact
with the cavity field, the joint state can be decomposed as
\begin{multline}
\rho(t)  = \sum_{\Lambda,J,M_J,\Lambda^{\prime },J^{\prime },M^{\prime
}_J}\sigma_{\Lambda,J,M_J,\Lambda^{\prime },J^{\prime },M^{\prime }_J}\Big(%
\ket{\Lambda,J,M_J}\bra{\Lambda',J',M'_J} \\
\otimes \ket{\alpha_{M_J}}\bra{\beta_{M'_J}}\Big)(t),%
\label{rho}
\end{multline}
where $\Lambda, J,M_J$ are labels corresponding to $\Lambda$ labeled irreps
with total angular momentum $J$ and $J^z$ projection $M_J$ (see \cite{Arrechi:05}), and $%
\sigma_{\Lambda,J,M_J,\Lambda^{\prime },J^{\prime },M^{\prime }_J}=\mbox{Tr}%
\lbrack \rho(0)\ket{\Lambda',J',M'_J}\bra{\Lambda,J,M_J}]$. The field states
$\ket{\alpha_{M_J}}$, $\ket{\beta_{M_J}}$, may depend on the angular
momentum projections. We describe evolution for the case where the field
states are Fock states for the single photon probe protocol and when they
are coherent states for the geometric phase gate. Depending on the protocol
used, the field states themselves may be entangled with the state of an
ancillary spin but we focus on computing fidelities for evolution steps
where the ancilla is non interacting.

Consider the evolution during an atom field coupling stage. The equation of
motion for the joint state is
\begin{align}
\dot{\rho}(t) & = \mathcal{L}(\rho(t)) \nonumber\\
& =  -i[V_Z,\rho(t)]+\frac{\kappa}{2} (2a \rho(t) a^{\dagger}-a^{\dagger} a
\rho(t)-\rho(t)a^{\dagger}a).\label{rhodot}
\end{align}
The evolution conserves the quantum numbers $\Lambda,J$ hence we can compute
the action $e^{\mathcal{L}t}A^{M_J,M^{\prime }_J}(0)$ where:
\[
A^{M_J,M^{\prime }_J}(t)\equiv \ket{\Lambda,J,M_J}\bra{\Lambda',J',M'_J}%
(t)\otimes \ket{\alpha_{M_J}}\bra{\beta_{M'_J}}(t).
\]
The solutions are easily verified to be given by
\begin{multline}
A^{M_J,M^{\prime }_J}(t) = \sum_{n=0}^\infty\frac{b^n_{M_JM^{\prime
}_J}(t)}{n!} e^{-(i2g_ZM_J+\kappa/2)a^\dagger a t} \\
\times a^nA^{M_J,M^{\prime }_J}(0){(a^\dagger)}^ne^{(i2g_ZM^{\prime
}_J-\kappa/2) a^\dagger a t}
\label{evolveA}
\end{multline}
where
\begin{equation}
b_{M_JM^{\prime }_J}(t)=\frac{\kappa\left(1-e^{-[\kappa+i2g_Z(M_J-M^{\prime
}_J)]t}\right)}{\kappa+i2g_Z(M_J-M^{\prime }_J)}.
\end{equation}
The evolved state is then
\[
\rho(t)=e^{\mathcal{L}t}\rho(0)=\sum_{\Lambda,J,M_J,\Lambda^{\prime
},J^{\prime },M^{\prime }_J}\sigma_{\Lambda,J,M_J,\Lambda^{\prime
},J^{\prime },M^{\prime }_J}A^{M_J,M^{\prime }_J}(t).
\]

In order to evaluate the performance of many body operations mediated by the
cavity we will calculate the process fidelity for implementing a many body
gate $U=e^{-i\chi S^z_{\mathcal{C}}}$ where $S^z_{\mathcal{C}}=\prod_{j\in
\mathcal{C}}\sigma^z_j$ on a configuration $\mathcal{C}$ of $m=|\mathcal{C}|$
spins. Many body measurements of the type $\langle S^z_{\mathcal{C}}\rangle$
are obtained using the many body gates with rotation angle $\chi=\pi/2$ as a
primitive.  The details are given in the following subsections with the main result
that the expected error scales like $\kappa/|g_Z|$.  

\subsection{Single photon mediated gate}

Consider a protocol where we begin with the separable state:
\begin{align*}
\rho(0) = & \rho_S(0)\otimes \rho_F(0) \\
= & \sum_{\Lambda,J,M_J,\Lambda^{\prime },J^{\prime },M^{\prime
}_J}\sigma_{\Lambda,J,M_J,\Lambda^{\prime },J^{\prime },M^{\prime }_J}
A^{M_J,M^{\prime }_J}(0)%
\end{align*}
where
\[
A^{M_J,M^{\prime }_J}(0)=\ket{\Lambda,J,M_J}\bra{\Lambda',J',M'_J}\otimes%
\ket{y_{\theta_+=\frac{\pi}{4}}}\bra{y_{\theta_+=\frac{\pi}{4}}},
\]
and the field states are defined
\[
\ket{y_{\theta_{+}}}=\cos\theta_+ \ket{0}+\sin\theta_+\ket{1},\quad %
\ket{y_{\theta_{-}}}=\ket{y_{\theta_{+}+\pi/2}}.
\]
Here the states $\ket{0},\ket{1}$ could denote photon number in a given mode
or a single photon in two orthonormal modes where only mode $\ket{1}$
interacts with the spins.

The discussion here can be generalized to the case that the cavity mode is
also entangled with an ancilla spin, with the mapping from $\left\{
\left\vert 0\right\rangle ,\left\vert 1\right\rangle \right\}  $ to $\left\{
\left\vert 0\right\rangle _{A}\otimes\left\vert \mathrm{vac}\right\rangle
,-\left\vert 1^{\prime}\right\rangle _{A}\otimes a^{\dag}\left\vert
\mathrm{vac}\right\rangle \right\}  $. In principle, the mapping does not hold
for the cavity decay, because the decay process without the ancilla spin (from
$\left\vert 1\right\rangle $ to $\left\vert 0\right\rangle $) cannot be mapped
to the decay process with an entangled ancilla (from $\left\vert 1^{\prime
}\right\rangle _{A}\otimes a^{\dag}\left\vert \mathrm{vac}\right\rangle $ to
$\left\vert 1^{\prime}\right\rangle _{A}\otimes\left\vert \mathrm{vac}%
\right\rangle $). However, if we are only interested in the fidelity of the
lattice spins, the analysis here is still valid, because the reduced density
matrices for the lattice spins (after tracing out the cavity mode and the
ancilla spin) are the same for both cases.

For atom field coupling over a period $t$, we have
from Eq. \ref{evolveA}
\begin{eqnarray*}
A^{M_J,M^{\prime }_J}(t) & = & \ket{\Lambda,J,M_J}\bra{\Lambda',J',M'_J}%
\otimes\frac{1}{2}\Big((1+b_{M_J,M^{\prime }_J}(t))\ket{0}\bra{0} \\
&  & +e^{(i2g_ZM^{\prime }_J-\kappa/2)t}\ket{0}\bra{1}+
e^{(-i2g_ZM_J-\kappa/2)t}\ket{1}\bra{0} \\
&  & +e^{(-i2g_Z (M_J-M^{\prime }_J)-\kappa)t}\ket{1}\bra{1}\Big).%
\end{eqnarray*}%
If we choose the interaction time $g_Z\tau=\frac{\pi}{2}$, then the action
on the state is
\begin{eqnarray}
\rho(\tau) & = & e^{\mathcal{L}\tau}\rho(0) \nonumber\\
& = & \frac{1}{2}\Big( \rho_S(0)\otimes\ket{0}\bra{0}+P(\rho_S(0))\otimes%
\ket{0}\bra{0} \nonumber\\
&  & \quad+e^{-\kappa \tau/2}\rho_S(0)e^{i\pi J^z}\otimes\ket{0}\bra{1} \nonumber\\
&  & \quad+e^{-\kappa \tau/2}e^{-i\pi J^z}\rho_S(0)\otimes\ket{1}\bra{0} \nonumber\\
&  & \quad+e^{-\kappa \tau}e^{-i\pi J^z}\rho_S(0)e^{i\pi J^z}\otimes\ket{1}\bra{1}%
\Big),
\label{imperfect}
\end{eqnarray}
where
\begin{multline*}
P(\rho_S(0)) = \sum_{\Lambda,J,M_J,\Lambda^{\prime },J^{\prime
},M^{\prime }_J}\sigma_{\Lambda,J,M_J,\Lambda^{\prime },J^{\prime
},M^{\prime }_J}b_{M_J,M^{\prime }_J}(t) \\
\times\ket{\Lambda,J,M_J}\bra{\Lambda',J',M'_J}.%
\end{multline*}
An ideal many body gate results if $\kappa=0$ and we measure the photon in a
rotated basis. Consider the case where the number of spins, $m$ is odd. If
we measure the photon in the basis $\ket{y_{\theta_{+}}}$, then we obtain
outcome $\pm 1$ with equal probabilities $p_{\pm}=\frac{1}{2}$ and the
resultant state is
\begin{eqnarray*}
\rho^{\pm}_S(\tau) & = & \frac{\mbox{Tr}\lbrack \rho(\tau) \ket{\theta_{\pm}}%
\bra{\theta_{\pm}}]}{\mbox{Tr}\lbrack \cdot]} \\
& = & e^{-i\theta_{\pm}(-i)^{m-1} S^z_{\mathcal{C}}}\rho_S(0)e^{i\theta_{%
\pm}(-i)^{m-1} S^z_{\mathcal{C}}}.%
\end{eqnarray*}
Say the target evolution operator is $e^{-i\theta_+ (-i)^{m-1} S^z_{\mathcal{%
C}}}$ but we obtain the measurement result $-1$. Such an outcome is
corrected for applying the locally generated unitary $S^z_{\mathcal{C}}$.
The case where $m$ is even is handled in a similar way but we measure the
photon in the basis
\[
\ket{x_{\theta_{+}}}=\cos\theta_+ \ket{0}+i\sin\theta_+\ket{1},\quad %
\ket{x_{\theta_{-}}}=-i\ket{x_{\theta_{+}+\pi/2}}.
\]
The measurement outcomes $\pm1$ are again equiprobable and the conditional
state is $\rho^{\pm}_S(\tau)=e^{-i\theta_{\pm}i^{m} S^z_{\mathcal{C}%
}}\rho_S(0)e^{i\theta_{\pm}i^{m} S^z_{\mathcal{C}}}$. The correction
procedure given the outcome $-1$ is as before.

We denote $\mathcal{E}$ the map for imperfect implementation of the gate
$U=e^{-i\chi S^z_{\mathcal{C}}}$ in the case of nonzero $\kappa$. For our
conditional protocol this can be represented as:
\begin{equation}
\mathcal{E}(\rho_S(0))=p_+ S_+(\rho_S(0))+p_- S^z_{\mathcal{C}%
}S_-(\rho_S(0))S^z_{\mathcal{C}},  \label{noisemap}
\end{equation}
where
\[
S_{\pm}(\rho_S(0))=\frac{1}{2p_{\pm}}\left(X(\theta_{\pm})\rho_S(0)X(%
\theta_{\pm})^{\dagger}+P(\rho_S(0)\right).
\]
We assume the product of local unitaries $S^z_{\mathcal{C}}$ can be
implemented with perfect fidelity.

For the remainder of this subsection, we fix $m$ odd and $m\bmod{4}=1$. The other
cases follow in a straightforward manner. From Eq. \ref{imperfect} we have
\[
X(\theta_{\pm})=\cos\theta_{\pm}\mathbf{1}+\sin\theta_{\pm}e^{-\kappa%
\tau/2}e^{-i\pi J^z}.
\]
Notice that now the probabilities for measuring the photon in $%
\ket{\theta_{\pm}}$ are not necessarily equal, rather one finds
\[
p_+-p_-=(1-e^{-\kappa \tau})\cos(2\theta_+).
\]
The process fidelity $F_{\mathrm{pro}}(\mathcal{E},U)$ measures how close a
quantum operation $\mathcal{E}$ is to the ideal operation $U$ as measured by
some suitable metric. The fidelity measure we use is the overlap between the
induced Jamio\l kowski-Choi state representations of the operations \cite%
{Nielsen:05}. The action of the unitary on a complete operator basis $\{%
\ket{\Lambda,J,M_J}\bra{\Lambda',J',M'_J}\}$ is multiplication by a
unimodular number:
\[
U\ket{\Lambda,J,M_J}\bra{\Lambda',J',M'_J}U^{\dagger}=V_{M_J,M^{\prime }_J}%
\ket{\Lambda,J,M_J}\bra{\Lambda',J',M'_J},\newline
\]
with
\[
V_{M_J,M^{\prime }_J}=(\cos\chi+\sin\chi e^{-i\pi M_J})(\cos\chi+\sin\chi
e^{i\pi M^{\prime }_J)}).
\]
Whereas the action of the map $\mathcal{E}$ on the same basis by
multiplication by a complex number:
\begin{multline*}
\mathcal{E}(\rho_S(0)) = \sum_{\Lambda,J,M_J,\Lambda^{\prime },J^{\prime
},M^{\prime }_J}\sigma_{\Lambda,J,M_J,\Lambda^{\prime },J^{\prime
},M^{\prime }_J}Q_{M_J,M^{\prime }_J} \\
\times\ket{\Lambda,J,M_J}\bra{\Lambda',J',M'_J},%
\end{multline*}%
where if we fix $\theta_+=\chi$ such that we approximate the unitary $U$
then
\begin{eqnarray*}
Q_{M_J,M^{\prime }_J} & = & \frac{1}{2}\Big[\cos^2\chi (1+b_{M_J,M^{\prime
}_J}(\tau))+\sin\chi\cos\chi e^{-\pi\kappa/4|g_Z|} \\
&  & \times(e^{-i\pi M_J}+ e^{i\pi M^{\prime }_J})+\sin^2\chi
e^{-\pi\kappa/2|g_Z|}e^{-i\pi (M_J-M^{\prime }_J)} \\
&  & +e^{-i\pi (M_J-M^{\prime }_J)}\big(\sin^2\chi(1+b_{M_J,M^{\prime
}_J}(\tau))-\sin\chi\cos\chi \\
&  & \times e^{-\pi\kappa/4|g_Z|}(e^{-i\pi M_J}+ e^{i\pi M^{\prime }_J})+\cos^2\chi
\\
&  & \times e^{-\pi\kappa/2|g_Z|}e^{-i\pi (M_J-M^{\prime }_J)}\big)\big)\Big].%
\end{eqnarray*}%
Of course for the case of no decay, $\kappa=0$, then $Q_{M_J,M^{\prime }_J}=V_{M_J,M^{\prime }_J}$.

The process fidelity is readily computed using the fact that the noise map $%
\mathcal{E}(\rho_S(0))$ commutes with the target unitary $U$. Hence, we can
compute the fidelity which measures how close the noisy map $\mathcal{E}%
^{\prime }(\rho_S(0))=U^{\dagger} \mathcal{E}(\rho_S(0)) U$ is to the ideal
operation, i.e. the identity operation:
\[
F_{\mathrm{pro}}(\mathcal{E},U)=F_{\mathrm{pro}}(\mathcal{E}^{\prime },%
\mathcal{I})=_{S,S^{\prime }}\langle \Phi^{+}|\rho_{\mathcal{E}^{\prime
}}|\Phi^{+}\rangle_{S,S^{\prime }}.
\]
Here we are computing the overlap of the Jamio\l kowski-Choi representations
of the maps as states in the Hilbert space $\mathcal{H}_S\otimes \mathcal{H}%
_{S^{\prime }}$ containing our system space and a copy each with dimension $%
D $:
\begin{align*}
|\Phi^+\rangle_{S,S^{\prime }}  &=  \frac{1}{\sqrt{D}}\sum_{\Lambda,J,M_J}%
\ket{\Lambda,J,M_J}_S\otimes \ket{\Lambda,J,M_J}_{S^{\prime }}, \\
\rho_{\mathcal{E}^{\prime }}  &=  \mathcal{I}_{S}\otimes\mathcal{E}^{\prime}%
_{S^{\prime }}(\ket{\Phi^+}_{S,S^{\prime }}) \\
 &=  \frac{1}{D}\sum_{\Lambda,J,M_J,\Lambda^{\prime },J^{\prime },M^{\prime
}_J}Q_{M_J,M^{\prime }_J}V^{\ast}_{M_J,M^{\prime }_J} \\
  &\phantom{=}\times \ket{\Lambda,J,M_J}_S \bra{\Lambda',J',M'_J}\otimes \ket{\Lambda,J,M_J}%
_{S^{\prime }} \bra{\Lambda',J',M'_J}.%
\end{align*}%
Hence
\[
F_{\mathrm{pro}}(\mathcal{E},U)=\frac{1}{D^2}\sum_{\Lambda,\Lambda^{\prime
}}\sum_{J,J^{\prime }}\sum_{M_J=-J}^J\sum_{M^{\prime }_J=-J^{\prime
}}^{J^{\prime }} R_{M_J,M^{\prime }_J},
\]
where
\[
R_{M_J,M^{\prime }_J}=Q_{M_J,M^{\prime }_J}V^{\ast}_{M_J,M^{\prime }_J}.
\]
Now $R_{M_J,M_J}=0$, and the for off diagonal elements, for $\kappa/|g_Z|\ll 1$ we find%
,
\begin{widetext}
\begin{equation}
\Re[R_{M_J,M'_J}]\approx
\begin{cases}
1-\frac{\pi}{2} \frac{\kappa}{2|g_Z|}+\frac{\pi^2}{4}(\frac{\kappa}{2g_Z})^2 & M_J-M'_J\ \rm{even} \\
1-\Big(\frac{\pi}{2}+
\frac{(-1)^{(2 M_J+1)/2}\sin(4\chi)}{2(M_J-M'_J)}\Big)\frac{\kappa}{2|g_Z|}+
\left(\frac{(-1)^{(2 M_J+1)/2}\sin(4\chi)}{4(M_J-M'_J)}+\frac{\cos(4\chi)+1}{2(M_J-M'_J)^2}+\frac{\pi^2(3+\cos(4\chi))}{16}\right)
(\frac{\kappa}{2g_Z})^2 & M_J-M'_J\  \rm{odd}.
\end{cases}
\label{expansiona}
\end{equation}
\end{widetext}
The above expression is made a bit simpler by counting $c^m_J$, the number
of inequivalent spin$-J$ irreps
of the $m$ fold symmetrized direct product of $SU(2)$, where $m$ is the
number of spin$-1/2$ particles in the system. The dimension can be computing
using Young tableau \cite{Sagan:90}:
\[
c^m_J=\frac{2J+1}{\frac{m}{2}+J+1} \left(%
\begin{array}{c}
m \\
\frac{m}{2}-J%
\end{array}%
\right).
\]
Making use of the fact that $Q_{M^{\prime }_J,M_J}=Q_{M_J,M^{\prime
}_J}^{\ast}$, the fidelity is
\[
F_{\mathrm{pro}}(\mathcal{E},U)=\frac{1}{2^{2m}}\sum_{J,J^{\prime
}=1/2}^{m/2}c^m_Jc^m_{J^{\prime }}\sum_{M_J=-J}^J\sum_{M^{\prime
}_J=-J^{\prime }}^{J^{\prime }} \Re[R_{M_J,M^{\prime }_J}].
\]
From Eq. \ref{expansiona} we find a lower bound for the fidelity:
\begin{equation}
F_{\mathrm{pro}}(\mathcal{E},U)> 1-\frac{\pi}{2} \frac{\kappa}{2|g_Z|}.
\label{lower1}
\end{equation}

\subsubsection{Protocol with a detector}

The above computation of fidelity may be overly pessimistic because one
could adopt a strategy where the output of the cavity is continuously
monitored for leakage of a photon during the coupling and altering the
protocol accordingly. Consider the situation where we have a perfect photon
detector outside the cavity that measures the presence of a leaked photon.
In the case of a null result, the system evolves via a non-Hermitian
Hamiltonian: $\rho(t)=\frac{e^{\mathcal{L}t}\rho(0)}{\mbox{Tr} \lbrack \cdot]%
}$ where $\mathcal{L}[\rho(t)]=-i(V_Z-i\frac{\kappa}{2} a^{\dagger}a)%
\rho(t)+i\rho(t)(V_Z+i\frac{\kappa}{2} a^{\dagger}a)$. The operator basis
elements then evolve as
\begin{eqnarray*}
A_{\mathrm{null}}^{M_J,M^{\prime }_J}(t) & = & \ket{\Lambda,J,M_J}%
\bra{\Lambda',J',M'_J}\otimes\frac{1}{1+e^{-\kappa t}}\Big(\ket{0}\bra{0} \\
&  & +e^{(i2g_ZM^{\prime }_J-\kappa/2)t}\ket{0}\bra{1}+
e^{(-i2g_ZM_J-\kappa/2)t}\ket{1}\bra{0} \\
&  & +e^{(-i2g_Z (M_J-M^{\prime }_J)-\kappa)t}\ket{1}\bra{1}\Big),%
\end{eqnarray*}%
whereas in the case of detected photon at time $t$ we have
\begin{equation}
\begin{array}{lll}
A_{\mathrm{det}}^{M_J,M^{\prime }_J}(t) & = & \ket{\Lambda,J,M_J}%
\bra{\Lambda',J',M'_J} e^{-i2g_Z (M_J-M^{\prime }_J))t}.%
\end{array}
\label{detA}
\end{equation}
The evolution in the later case is unitary and we can restore the system to
its prior state by applying the locally generated unitary operator
$W_{\mathrm{corr}}(t)=e^{i2g_Z t J^z}$. We adopt the following protocol to
implement $U=e^{-i\chi S^z_{\mathcal{C}}}$:

\begin{itemize}
\item Prepare the ancilla as in Sec. \ref{singlephot} and let the photon
interact with the system for a time $\tau=\pi/2g_Z$.

\item If no photon is detected, then measure the photon inside the cavity in
the rotated basis $\ket{\chi_{\pm}}$ and if the outcome is $\ket{\chi_-}$
apply local gate $S^z_{\mathcal{C}}$ as in Sec. \ref{singlephot}. End.

\item If a photon is detected at time $t$ apply the local correction gate $%
W_{\mathrm{corr}}(t)$. Repeat.
\end{itemize}

In the above protocol, the unitary $U$ is approximated after a sequence of $%
m-1$ clicks and a final null count, an event which occurs with probability:
\begin{align*}
p_m & =  \int _0^{\tau}dt_m p_{\mathrm{null}}(t_m)\int_0^{\tau}\cdots \int
_0^{\tau}dt_{m-1}\ldots dt_{1} p_{\mathrm{det}}(t_m) \\
& =  (\frac{|g_Z|}{\pi \kappa})^m (\frac{\pi\kappa}{2|g_Z|}-e^{\pi
\kappa/2|g_Z|}+1)(\frac{\pi\kappa}{2|g_Z|}+e^{\pi \kappa/2|g_Z|}-1)^{m-1},%
\end{align*}%
where the integration measure is $dt_k=1/\tau$ and
\[
p_{\mathrm{null}}(t_k)=\frac{1+e^{-\kappa t_k}}{2},\quad p_{\mathrm{det}%
}(t_k)=1-p_{\mathrm{null}}(t_k).
\]
Assuming for simplicity that the total time to detect a photon, apply the
correction gate and re-prepare a probe photon is roughly $\tau$, then the
mean $\bar{t}_{\mathrm{gate}}$ and variance $\Delta t_{\mathrm{gate}}$ of
the time to perform the gate is
\begin{align*}
\bar{t}_{\mathrm{gate}} & =  \tau \sum_{m=1}^{\infty} m p_m \\
& =  \frac{\tau e^{\pi \kappa/2|g_Z|}\pi\kappa/|g_Z|}{e^{\pi
\kappa/2|g_Z|}(\pi\kappa/2|g_Z|+1)-1}\leq 2\tau, \\
\Delta t_{\mathrm{gate}} & =  \sqrt{\bar{t^2}_{\mathrm{gate}}-\bar{t}_{%
\mathrm{gate}}^2} \\
& =  \frac{\tau \sqrt{e^{\pi \kappa/2|g_Z|} (e^{\pi \kappa/2 |g_Z|}(\pi
\kappa/2|g_Z|-1)+1)\pi\kappa/|g_Z|}}{e^{\pi \kappa/2|g_Z|}(\pi\kappa/2|g_Z|+1)-1}\\
&\leq
\sqrt{2}\tau.&%
\end{align*}
Notice that these values are bounded above.

The fidelity is easily calculated by noting that the action of each map
associated with a detector click is a unitary operation which is undone by a
correction step. So the only step in the protocol that acts non trivially is
the step with the final null count. Evaluating the fidelity as before but
using the expression $A_{\mathrm{null}}^{M_J,M^{\prime }_J}(t)$ we find
\begin{align*}
F_{\mathrm{pro}}(\mathcal{E},U) & =  \frac{1}{2} \left(1+\cos^2(2\chi)+%
\frac{2e^{\pi \kappa/4g_Z}}{1+e^{\pi\kappa/2g_Z}}\sin^2(2\chi)\right) \\
& =  1-\frac{\pi^2\kappa^2}{64 g_Z^2}\sin^2(2\chi)+O((\kappa/2g_Z)^4),%
\end{align*}%
where the expansion is valid for $\kappa/|g_Z|\ll 1$.

In fact we can do better. Using our knowledge of the necessary gate time $%
\tau$ in the event of a null detection it is advantageous to prepare the
initial photon probe state
\[
(\ket{0}+e^{\pi \kappa/4|g_Z|}\ket{1})/\sqrt{1+e^{\pi \kappa/2|g_Z|}}.
\]
Evolution of the basis states is
\begin{eqnarray*}
A_{\mathrm{null}}^{M_J,M^{\prime }_J}(t) & = & \ket{\Lambda,J,M_J}%
\bra{\Lambda',J',M'_J}\otimes\frac{1}{1+e^{\pi\kappa/2|g_Z|-\kappa t}}\Big(%
\ket{0}\bra{0} \\
&  & +e^{(i2g_ZM^{\prime }_Jt+\pi\kappa/4|g_Z|-\kappa t/2)}\ket{0}\bra{1} \\
&  & +e^{(-i2g_ZM_Jt+\pi\kappa/4|g_Z|-\kappa t/2)}\ket{1}\bra{0} \\
&  & +e^{(-i2g_Z (M_J-M^{\prime }_J)t+\pi\kappa/2|g_Z|-\kappa t)}\ket{1}\bra{1}%
\Big),%
\end{eqnarray*}%
and $A_{\mathrm{det}}^{M_J,M^{\prime }_J}(t)$ is as in Eq. \ref{detA}. After
a time $\tau$ a null detection gives the target evolution and the fidelity
is one. As before, in the case of a photodetection event, the system
evolution can be reversed and the protocol repeated.

Now the probabilities for a null count or detection of a photon are modified
to:
\[
p_{\mathrm{null}}(t_k)=\frac{1+e^{\pi\kappa/2|g_Z|-\kappa t_k}}{%
1+e^{\pi\kappa/2|g_Z|}},\quad p_{\mathrm{det}}(t_k)=1-p_{\mathrm{null}}(t_k).
\]
The mean $\bar{t}_{\mathrm{gate}}$ and variance $\Delta t_{\mathrm{gate}}$
of the time to perform the gate is
\begin{eqnarray*}
\bar{t}_{\mathrm{gate}} & = & \tau \sum_{m=1}^{\infty} m p_m \\
& = & \frac{\tau (1+e^{\pi \kappa/2|g_Z|})\pi\kappa/2|g_Z|}{\pi\kappa/2|g_Z|+e^{%
\pi \kappa/2|g_Z|}-1}, \\
\Delta t_{\mathrm{gate}} & = & \sqrt{\bar{t^2}_{\mathrm{gate}}-\bar{t}_{%
\mathrm{gate}}^2} \\
& = & \frac{\tau \sqrt{(1+e^{\pi \kappa/2|g_Z|}) (e^{\pi \kappa/2|g_Z|}(\pi
\kappa/2g_Z-1)+1)\pi\kappa/2|g_Z|}}{\pi\kappa/2|g_Z|+e^{\pi \kappa/2|g_Z|}-1}.%
\end{eqnarray*}%
The mean and variance of the gate time is now unbounded with increasing $%
\kappa$ but for $\kappa/|g_Z|\ll 1$ the values are comparable to the prior
case with the photon probe prepared in $(\ket{0}+\ket{1})/\sqrt{2}$.

For a situtation with finite detector efficiency $\eta$ which can be modeled
as a rank $2$ projector on the photon, the fidelity will degrade ultimately
to the case of no detector as derived above.

\subsection{Geometric phase gate}

\label{gpfid}

In order to evaluate the effect of cavity decay during the the geometric
phase gate, we are particularly interested in the case where initially $%
A^{M_J,M^{\prime }_J}(0)=\ket{\Lambda,J,M_J}\bra{\Lambda',J',M'_J}%
\otimes|\alpha_{M_J}\rangle\langle\beta_{M^{\prime }_J}|$, with $%
|\alpha_{M_J}\rangle$, $\ket{\beta_{M'_J}}$ coherent states. This kind of
factorization is true at any stage of spin coupling to the field. Using, Eq. %
\ref{evolveA}, the sum becomes an exponential and the evolved state is
\begin{align}
\rho(t) & = e^{\mathcal{L}t}\rho(0) \nonumber\\
& = \sum_{\Lambda,J,M_J,\Lambda^{\prime },J^{\prime },M^{\prime
}_J}e^{d(t)}\sigma_{\Lambda,J,M_J,\Lambda^{\prime },J^{\prime },M^{\prime
}_J}\ket{\Lambda,J,M_J}\bra{\Lambda',J',M'_J} \nonumber\\
& \hspace{0.5in}  \otimes |e^{-(igM+\kappa/2)t}\alpha_{M_J}\rangle\langle
e^{-(igM^{\prime }+\kappa/2)t}\beta_{M^{\prime }_J}|,%
\end{align}
\label{evolved}
where
\begin{equation}
d(t)=\alpha_{M_J}\beta_{M^{\prime }_J}^*b_{M_J,M^{\prime }_J}(t)
-(|\alpha_{M_J}|^2+|\beta_{M^{\prime }_J}|^2){\textstyle\frac{1-e^{-\kappa t}%
}{2}}.
\end{equation}
For completeness, an alternate derivation of the evolution using the
characteristic equation for the joint state is give in Appendix \ref{AppA}.

We ignore decay during the displacement stages of the evolution (i.e. we
assume these are done quickly relative to the decay rate), and we assume
that the system particles do not interact with the field during these steps.
In order to perform logical operations on the protected momory, we do not
need an ancilla and there are seven time steps beginning with the cavity in
the vacuum state:
\[
D(-\beta)e^{-i \tau_5 V_Z}D(-\alpha)e^{i \tau_3 V_Z}D(\beta)e^{-i \tau_1
V_Z}D(\alpha),
\]
as described in Sec. \ref{gpimp}. Let $\tau_5=\tau_3=\tau_1$ so that the
periods of spin field coupling are all equal in duration. Notice the change
in sign of the evolution during the period $\tau_3$. This can be
accommodated by changing the sign of the coupling parameter $g_Z$ by e.g.
changing the sign of field detuning. In order that the field state return to
the vacuum at the end of the sequence, we choose $\alpha^{\prime -\kappa
\tau_1},\beta^{\prime -\kappa \tau_1}$. The total sequence then yields the
output state:

\begin{multline*}
\rho_{\mathrm{out}}  =  \displaystyle{\sum_{\Lambda,J,M_J,\Lambda^{\prime
},J^{\prime },M^{\prime }_J}} \sigma_{\Lambda,J,M_J,\Lambda^{\prime
},J^{\prime },M^{\prime }_J}R_{M_J,M^{\prime }_J} \\
\times e^{i\chi\sin(\phi+2g_Z\tau_1 M_J)}\ket{\Lambda,J,M_J}%
\bra{\Lambda',J',M'_J} \\
\times e^{-i\chi\sin(\phi+2g_Z\tau_1 M^{\prime }_J)}\otimes \ket{0}\bra{0},%
\end{multline*}
where we defined $R_{M_J,M^{\prime }_J}=e^{d(t_2)+d(t_4)+d(t_6)}$ and $%
\chi=|\alpha \beta |(e^{-3\kappa \tau_1/2}+e^{-\kappa \tau_1/2})/2$. This can be interpreted as coherent evolution 
with an effective evolution operator
\[
e^{-iH_{\rm int}T}=e^{i\chi\sin(\phi+2g_Z\tau_1 J^z)},
\]
where $T$ is an effective time for the gate, followed by dephasing in the $%
\{\Lambda,J,M_J\}$ basis. Matrix elements diagonal in $M_J$ are invariant.
For $m$ even, the parameters $g_Z \tau_1=\pi/2$, $|\alpha|=|\beta|$, $%
\phi=\pm\pi/2$, generate
\[
U=\exp[\mp i(-1)^{\frac{m}{2}} |\alpha|^2S^z_{\mathcal{C}} (e^{-3\kappa
\tau_1/2}+e^{-\kappa \tau_1/2})/2].
\]
The strength of the dephasing and decay is then,
\begin{widetext}
\begin{multline}
R_{M_J,M'_J}=
\exp\Big[\frac{(M_J-M'_J)|\alpha|^2e^{-\pi\kappa/|g_Z|}(1+e^{\pi\kappa/2|g_Z|})(2(1-e^{\pi \kappa/2|g_Z|})(M_J-M'_J)\pm ((-1)^{M_J}-(-1)^{M'_J})e^{\pi\kappa/4|g_Z|}\kappa/2|g_Z|)}{(M_J-M'_J)^2+(\kappa/2g_Z)^2}\Big]\\
\times\exp\Big[\mp i \frac{((-1)^{M_J}-(-1)^{M'_J})|\alpha|^2e^{-3\pi\kappa/4|g_Z|}(1+e^{\pi \kappa/2|g_Z|})(\kappa/2g_Z)^2}{(M_J-M'_J)^2+(\kappa/2g_Z)^2}\Big].
\label{deph}
\end{multline}
Note that $R_{M'_J,M_J}=R_{M_J,M'_J}^{\ast}$.  For $\kappa/|g_Z|\ll 1$ and $M_J\neq M'_J$,

\begin{equation}
\Re[R_{M_J,M'_J}]\approx \begin{cases}1-4\pi|\alpha|^2\frac{\kappa}{2|g_Z|}+4\pi^2(\frac{\kappa}{2g_Z})^2|\alpha|^2(1+2|\alpha|^2) & M_J-M'_J\  \rm{even} \\
1-4\pi|\alpha|^2\big(1+\frac{1}{M_J-M'_J}\big)\frac{\kappa}{2|g_Z|}+\frac{4|\alpha|^2((M_J-M'_J)\pi-1)((M_J-M'_J)(2\pi|\alpha|^2+\pi)-2|\alpha|^2)}{(M_J-M'_J)^2}(\frac{\kappa}{2g_Z})^2 & M_J-M'_J\  \rm{odd}. \end{cases}
\label{expansionb}
\end{equation}
\end{widetext}
For $m$ odd, the parameters $g_Z \tau_1=\pi/2$, $|\alpha|=|\beta|$, $%
\phi=0(\pi)$, generate
\[
U=\exp[\mp i(-1)^{\frac{m-1}{2}} |\alpha|^2S^z_{\mathcal{C}} (e^{-3\kappa
\tau_1/2}+e^{-\kappa \tau_1/2})/2].
\]
The strength of the dephasing is the same as Eq. \ref{deph} but with the
replacements $M_J\rightarrow (2M_J-1)/2$ and $M^{\prime }_J\rightarrow
(2M^{\prime }_J-1)/2$. 


As in the case of the single photon mediated gate, the noisy implementation
of the geometric phase gate commutes with the target unitary $U=e^{-i\chi
S^z_{\mathcal{C}}}$. The process fidelity is then,
\[
F_{\mathrm{pro}}(\mathcal{E},U)=\frac{1}{2^{2m}}\sum_{J,J^{\prime}}^{m/2}c^m_Jc^m_{J^{\prime }}\sum_{M_J=-J}^J\sum_{M^{\prime
}_J=-J^{\prime }}^{J^{\prime }} \Re[R_{M_J,M^{\prime }_J}],
\]
where we make use of the fact that $R_{M^{\prime }_J,M_J}=R_{M_J,M^{\prime
}_J}^{\ast}$. The magnitude of the coherent state amplitude is chosen to
best approximate $U$ by fixing $|\alpha|=|\beta|$ and:
\[
\chi=|\alpha|^2(e^{-3\pi\kappa/4|g_Z|}+e^{-\pi\kappa/4|g_Z|})/2.
\]
It is quickly verified that for $\kappa=0$, $F_{\mathrm{pro}}(\mathcal{E}%
,U)=1$. By Eq. \ref{expansionb} we find the lower bound
\begin{equation}
\begin{array}{lll}
F_{\mathrm{pro}}(\mathcal{E},U)&\geq&1-\frac{4\pi\chi\kappa/|g_Z|}{%
(e^{-3\pi\kappa/4|g_Z|}+e^{-\pi\kappa/4|g_Z|})}\\
&>& 1-\frac{4\pi \chi\kappa}{%
2|g_Z|} \Big(1+\frac{\pi\kappa}{2|g_Z|}\Big). 
\end{array}
 \label{lower2}
\end{equation}

\begin{figure}[tbp]
\begin{center}
\includegraphics[width=\columnwidth]{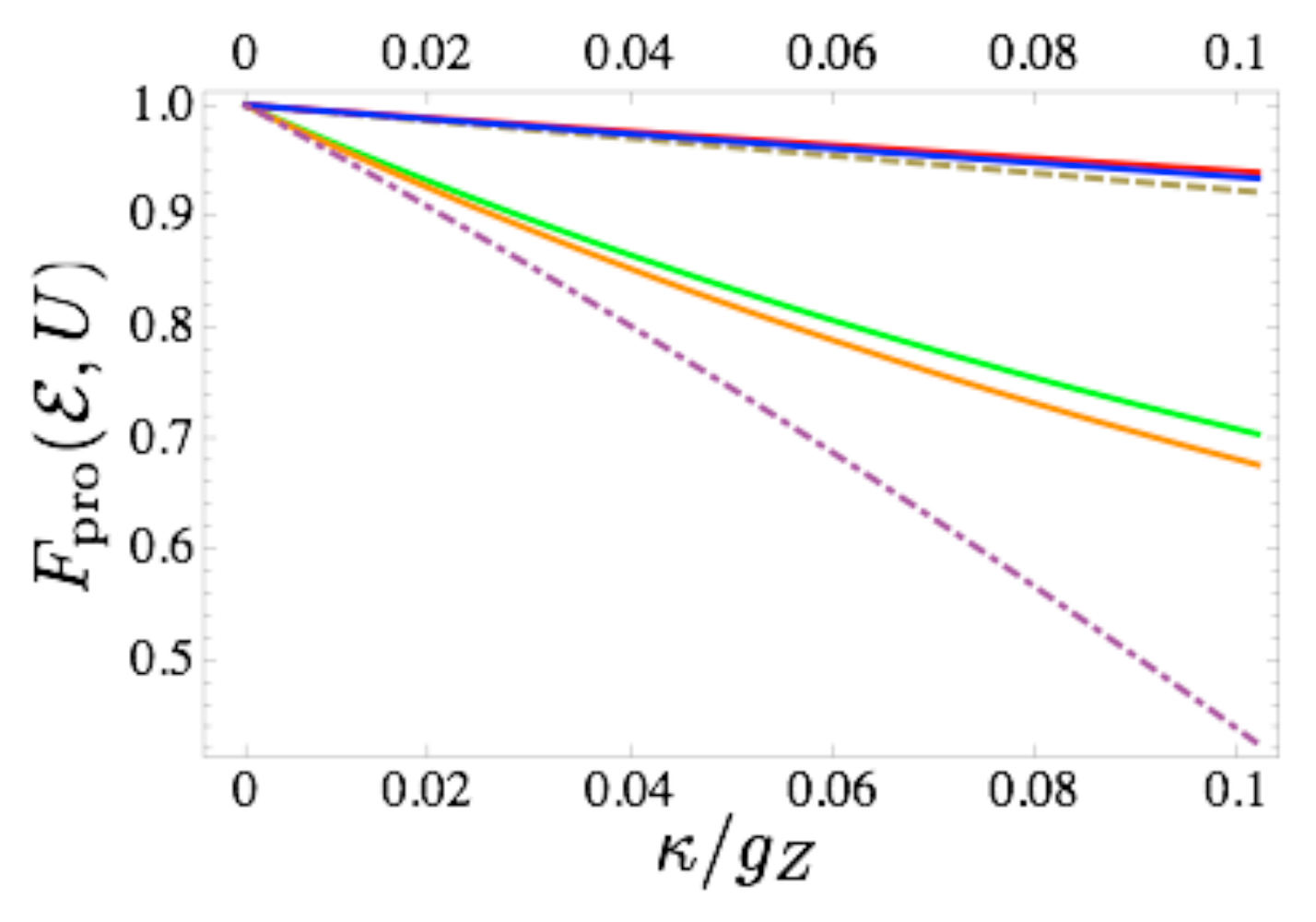}
\end{center}
\caption{Process fidelity of an implementation $\mathcal{E}$ of the many
body gate $U=e^{-i\frac{\protect\pi}{4}\prod_{j=1}^m \protect\sigma^z_j}$ as
a function of cavity decay $\protect\kappa$ (in units of the particle field
coupling strength $g_Z$). Here the time is chosen so that $g_z \protect\tau=%
\protect\pi/2$ where $\protect\tau$ is the time spent during each stage of
coherent coupling between field and particles. Plots are shown for an
implementation using the geometric phase gate on $m=9$ spins (green) and $%
m=25$ spins (orange); and using a single photon for $m=9$ (red), $m=25$
(blue). Also shown are the lower bounds on fidelity from Eq. \protect\ref%
{lower1} (dashed) and Eq. \protect\ref{lower2} (dot-dashed). For the phase
gate, we choose the coherent state amplitudes according to $|\protect\alpha%
\protect\beta|(e^{-3\protect\kappa \protect\tau/2}+e^{-\protect\kappa
\protect\tau/2})/2=\frac{\protect\pi}{4}$. It is assumed that no decay
occurs during the field displacement stages.}
\label{fig:6}
\end{figure}

In closing, note that the process fidelity for the case in which the target
evolution is unitary can be related to the average fidelity via \cite%
{Nielsen:05}:
\begin{align*}
F_{\mathrm{ave}}(\mathcal{E},U) & = \int_{\ket{\psi}\in \mathcal{H}_S}
d\psi F(\mathcal{E}(\ket{\psi}),U\ket{\psi}\bra{\psi}U^{\dagger})) \\
& = \frac{F_{\mathrm{pro}}(\mathcal{E},U) D+1}{D+1}.%
\end{align*}
Topologically ordered states have the property that for pure states, when
the system is divided into two connected domains, the subsystem entropy
scales like the size of the boundary \cite{Hamma:05}. For the surface codes
and in the case where the
two subsystems are just one string of spins and the rest, this implies that
the subsystem entropy of the string is nearly maximal because by isotopy
the state of any string on the lattice can be deformed to any other string
in the same homology equivalence class.  Hence we expect that the subsystem of spins that are
acted on during the gate has equal weight on most states in its Hilbert
space and the measure of fidelity as an average measure over pure states is
a good one.

\section{Fidelity with other decoherence mechanisms}\label{Sec:OtherDec}

Up to now we have ignored decoherence mechanisms such as radiative decay of
the spins into all modes of the electromagnetic field, and possible sources
of noise such as fluctuating optical trapping fields and stray magnetic
fields. Many of these effects will be system dependent, however we can make
some quantitative statements for the case that the noise is isotropic. This
is a reasonable working assumption because in order to obtain the spin
lattice models used for protected quantum memories, it is assumed that the
qubit levels are degenerate. Hence absent any special symmetry imposed on
the environment and control fields, we expect the noise and radiative decay
to act isotropically on the spins.

\subsection{Collective depolarization}

In the case where the decoherence channels correspond to environmental modes
that couple coherently to all the spins, we can describe the system as
undergoing collective decoherence. This would be case, e.g. for trapped
polar molecules where the transition microwave wavelength is much larger
than the optical wavelength spacing between molecules. The map describing
collective depolarization is described by applying a the same random $SU(2)$
rotation to all qubits in the system
\begin{align*}
\mathcal{E}_{\mathrm{cd}}(\rho) & = (1-p)\rho+p\int_{\Omega\in
SU(2)}d\Omega [U(\Omega)]^{\otimes m}\rho [U^{\dagger}(\Omega)]^{\otimes m}\\
& = (1-p)\rho +p\sum_{J,\Lambda,\Lambda^{\prime
}}[\rho^J_{\Lambda,\Lambda^{\prime }}]\otimes \frac{\mathbf{1}_{2J+1}}{2J+1},%
\end{align*}
where $\rho^J_{\Lambda,\Lambda^{\prime }}=\sum_{M_J}\langle
\Lambda,J,M_J|\rho|\Lambda^{\prime },J,M_J\rangle$. The strength of the
collective depolarization is parameterized by $p$ such that for isotropic
decay at rate $\gamma$, over a time period $t$, $p=1-e^{-\gamma t}$.
Essentially the collective depolarization erases coherences between
different $J$ quantum numbers and maximally mixes the reduced state within
each $J$ block. Accordingly, this map commutes with the map $\mathcal{E}_g$
corresponding to the cavity decay derived above: $\mathcal{E}_{\mathrm{cd}%
}\circ \mathcal{E}_g(\rho)=\mathcal{E}_g\circ \mathcal{E}_{\mathrm{cd}%
}(\rho) $, and as before we can compute the process fidelity for a process
with cavity decay and collective depolarization as:
\begin{equation}
F_{\mathrm{pro}} (\mathcal{E}_{\mathrm{cd}}\circ \mathcal{E}_g,U)=
_{S,S^{\prime }}\langle \Phi^{+}|\rho_{\mathcal{E}^{\prime
}}|\Phi^{+}\rangle_{S,S^{\prime }},
\end{equation}
where $\mathcal{E^{\prime }}(\rho)=U^{\dagger}(\mathcal{E}_{\mathrm{cd}%
}\circ \mathcal{E}_g (\rho)) U$ and
\begin{multline*}
\rho_{\mathcal{E^{\prime }}} = \frac{1}{2^{m}}\sum_{\Lambda,J,M_J,%
\Lambda^{\prime },J^{\prime },M^{\prime }_J}\ket{\Lambda,J,M_J}_S%
\bra{\Lambda',J',M'_J}\\
\otimes((1-p)R_{M_J,M^{\prime }_J}\ket{\Lambda,J,M_J}_{S^{\prime }}%
\bra{\Lambda',J',M'_J} \\
+\frac{p\delta_{M_J,M^{\prime }_J}\delta_{J,J^{\prime }}}{2J+1}%
\sum_{M^{\prime \prime }_J=-J}^{J}\ket{\Lambda,J,M''_J}_{S^{\prime }}%
\bra{\Lambda',J,M''_J}).%
\end{multline*}
We find
\begin{multline*}
F_{\mathrm{pro}} (\mathcal{E}_{\mathrm{cd}}\circ \mathcal{E}_g,U) =
(1-p)F_{\mathrm{pro}}(\mathcal{E}_g,U)\\
+\frac{p}{2^{2m}}\sum_{J=(1-(-1)^m)/4}^{m/2}(c^m_J)^2.%
\end{multline*}
The sum can be evaluated in closed form in terms of hypergeometric functions
and it quickly decays to zero, e.g. for $m$ large the second term scales like $\frac{p}{5}(\frac{m}{2})^{-3/2}$.
.

\subsection{Independent depolarization}

For environments with a correlation length small compared to the lattice spacing, as in the case of optical scattering by trapped atoms in an optical lattice we can model the decoherence as independent isotropic noise on each qubit. Under this assumption, the evolution equation will contain an additional Liouvillian
\[
\dot\rho=\frac{\gamma_{\rm eff}}{4}\sum_{k=1}^m\sum_{\alpha}(\sigma^{\alpha}_k\rho\sigma^{\alpha}_k-\rho),
\]
where the effective decay rate per particle is $\gamma_{\rm eff}=\gamma\, \bar{n}\, g^2/\Delta^2$ where $\gamma$ is the spontaneous decay rate and $\bar{n}$ is the average photon number in the cavity. For the single photon mediated gate and the geometric phase gate we can assume $\bar{n}\sim 1$ and $\bar{n}=|\alpha|^2$ respectively. The corresponding noise map is now
\[
\mathcal{E}_{\rm id}(\rho)=(1-m\,p)\rho +p\sum_{k=1}^m \tr_k[\rho]\otimes \frac{{\bf 1}_2}{2},
\]
where $\tr_k$ is the trace over the $k-$th spin and $p=1-e^{-\gamma_{\rm eff} t}$. Different from the case of collective depolarization, the operations $\mathcal{E}_g$ and $\mathcal{E}_{\rm id}$ do not commute, as can easily be easily verified on a two qubit system.

An estimate for the effect of spontaneous emission can still be obtained in the perturbative limit, where $m\,p\simeq m\,\gamma_{\rm eff}\tau\ll1$ for a gate time $\tau$. For both gates this time is $\tau\sim \pi\Delta/g^2$. The final state can then be approximated by
\[
\rho_{\rm out}\simeq \mathcal{E}_{\rm id}\circ\mathcal{E}_{\rm g}(\rho)
\]
and the corresponding process fidelity can be lower bounded by $F_{\rm pro}(\mathcal{E}_{\rm id}\circ\mathcal{E}_{\rm g},U)\geq (1-m\,p)F_{\rm pro}(\mathcal{E}_{\rm g},U)$.

\section{Conclusions}

We have analyzed the performance of cavity mediated many body gates
in a spin lattice.  To summerize the requirements for robust gates we require:
 high quality cavities, i.e. low loss rates $(\kappa/|g_Z|\ll 1$), 
 and dispersive coupling $\gamma,g\ll \Delta$.  These can be satisfied
in the strong coupling limit where $\gamma,\kappa\ll 1$.
Note that it is possible to have $\kappa\ll g < \gamma$ and still satisfy these requirements.
 Recent experiments \cite{Rempe:07} reported 3D trapping of Rb
atoms in a high finesse optical cavity with coupling parameters $%
(g,\gamma,\kappa)/2\pi=(16,3,1.4)$ MHz. Microwave cavities offer the
possibility of even better numbers. For example, superconducting strip line
cavities resonant at microwave transitions frequencies have been built \cite%
{Schoelkopf} with parameters $(g,\kappa)/2\pi=(200,0.1)$ MHz. These cavities
can be used to trap polar molecules and coherently control them on the
microwave transitions between rotational levels (the linewidths on such
excited rotational states are negligible)\cite{Zoller:06}. A difficulty
here may be interacting with the atoms using lasers in the vicinity of the
strip line cavities. One might try to trap without lasers using self
assembly with static electric fields but care should be taken to ensure that
the underlying lattice model is compatible with the setup.  Ultimately, 
we expect the idea of using quantum probes for many body control will
suggest new strategies for
 information processing in strongly correlated states of matter.

\section{Acknowledgements}

We gratefully acknowledge conversations with D. Bacon, H. P. Buchler, E.
Demler, A. V. Gorshkov, M. Hafezi, L. Ioffe. Work at Harvard is supported by
NSF, ARO-MURI, CUA, DARPA, AFOSR, and the Packard Foundation. Work at
Innsbruck is supported by the Austrian Science Foundation, the EU under
grants OLAQUI, SCALA, and the Institute for Quantum Information. %

\appendix

\section{A subsystem code}\label{AppB}

The Hamiltonian $H_{\rm cp}$ differs from the 2D Ising type model introduced in \cite%
{Bacon:06}, namely: $H^{\prime }=-J(\sum_{\mathrm{x-links}}\sigma ^{x}\sigma
^{x}+\sum_{\mathrm{y-links}}(\sigma ^{x}\sigma ^{x}+\sigma ^{z}\sigma
^{z})+\sum_{\mathrm{z-links}}\sigma ^{z}\sigma ^{z})$. However, both models
possess the same subsystem structure.
Stabilizer operators are generated by the $2(n-1)$ members of the set $%
\{V_{j}^{X},V_{j}^{Z}\}_{j=1}^{n-1}$ where the generators are adjacent
planes of $\sigma ^{x}$ operators in the $\hat{x}-\hat{y}$ plane and $\sigma
^{z}$ operators in the $\hat{y}-\hat{z}$ plane:
\[
V_{i}^{X}=\prod_{j,k=1}^{n}\sigma _{i,j,k}^{x}\sigma _{i+1,j,k}^{x},\quad
V_{k}^{Z}=\prod_{i,j=1}^{n}\sigma _{i,j,k}^{z}\sigma _{i,j,k+1}^{z}.
\]

The Hamiltonian $H_{\mathrm{cp}}$ can encode one qubit of information in a
subsystem of the total Hilbert space $\mathcal{H}_2^{\otimes n^3}$. For
clarity, we recall the argument given in \cite{Bacon:06} for the subsystem
structure of the energy eigenspaces. It is understood by considering
invariant subspaces of the Hamiltonian with respect to three sets of
operators: $\mathcal{L},\mathcal{V},\mathcal{T}$. The set $\mathcal{T}$,
which is a group, consists of all products of Pauli operators consisting of
an even number of $\sigma^x$ operators in each $\hat{x}-\hat{y}$ plane and
an even number of $\sigma^z$ operators in each $\hat{y}-\hat{z}$ plane. It
is generated under multiplication by the summands in the Hamiltonian $H$,
i.e.
\begin{equation}
\mathcal{T}=\langle
\{\sigma^x_{i,j,k}\sigma^x_{i+1,j,k},\sigma^x_{i,j,k}\sigma^x_{i,j+1,k},%
\sigma^z_{i,j,k}\sigma^z_{i,j+1,k},\sigma^z_{i,j,k}\sigma^z_{i,j,k+1}\}%
\rangle
\end{equation}
The Hamiltonian, in particular, is in the real span of $\mathcal{T}$. The
stabilizer set $\mathcal{V}$, also a group, is generated under
multiplication as
\begin{equation}
\mathcal{V}=\langle \{V^X_k,V^Z_k\}_{k=1}^{n-1}\rangle
\end{equation}
$\mathcal{V}$ is an abelian subgroup of $\mathcal{T}$. Finally, the set $%
\mathcal{L}$ consists of operators with an odd number of $\hat{y}-\hat{z}$
plane operators $L^X_i=\prod_{j,k=1}^n\sigma^x_{i,j,k}$ and an odd number of
$\hat{x}-\hat{y}$ plane operators $L^Z_k=\prod_{i,j=1}^n\sigma^z_{i,j,k}$,
i.e.
\begin{equation}
\mathcal{L}=\langle\{L^X_k\}\rangle/\langle\{V^X_k\}\rangle\ \widehat{}\
\langle\{V^Z_i\}\rangle/\langle\{V^Z_i\}\rangle
\end{equation}
This set is clearly not a group (e.g. it has no identity element) but $%
\mathcal{L}\cup\mathcal{V}$ is. Note that $\forall t\in \mathcal{T},v\in%
\mathcal{V},\ell \in \mathcal{L}$, the following commutation relations hold $%
vtv^{-1}=t,t\ell t^{-1}=\ell$.

We can partition the Hilbert space into $\pm 1$ eigenspaces of the $2(n-1)$
independent stabilizer generators $\{V^X_k,V^Z_i\}$:
\begin{equation}
\mathcal{H}=\oplus_{v^X,v^Z}\mathcal{H}_{v^X,v^Z}
\end{equation}
where $v^X=(v^X_1,v^X_2,\ldots, v^X_{n-1})$ is an $n-1$ bit string of the
eigenvalues of $V^X_k$ and $v^Z=(v^Z_1,v^Z_2,\ldots, v^Z_{n-1})$ is an $n-1$
bit string of the eigenvalues of $V^Z_k$. Because the Hamiltonian $H_{%
\mathrm{cp}}\in \mathrm{span}_{\mathbb{R}}\mathcal{T}$, its eigenspaces are
block diagonal in $\{v^X,v^Z\}$. Furthermore, because all elements of $%
\mathcal{L}$ commute with elements of $\mathcal{T}$, we further decompose
the eigenspaces as
\begin{equation}
\mathcal{H}_{v^X,v^Z}=\mathcal{H}_{v^X,v^Z}^{\mathcal{T}}\otimes \mathcal{H}%
_{v^X,v^Z}^{\mathcal{L}}
\end{equation}
Operators in $\mathcal{L}$ commute with $\mathcal{T}$ and $\mathcal{V}$ so
they leave those spaces invariant. In a given stabilizer eigenspace, any
operator in $\mathcal{L}$ can be reduced to the simple product of one plane
operator $L^X_1$ and one plane operator $L^Z_1$. Because $n$ is odd these
operators anticommute: $\{L^X_1,L^Z_1\}=0$, hence they form a representation
of a two dimensional Clifford algebra. By dimension counting then: $\mathrm{%
dim}\mathcal{H}_{v^X,v^Z}^{\mathcal{T}}=2^{n^3-2n+1}$ and $\mathrm{dim}%
\mathcal{H}_{v^X,v^Z}^{\mathcal{L}}=2$. It is in the subspace $\mathcal{L}$
that a logical qubit can be stored. Furthermore, the logical operators on
the qubit subspace correspond to single plane operators. Writing $%
L^Z_1L^X_1=iL^Y_1$, the algebra $\mathrm{span}_\mathbb{R}\{L^X_1,L^Y_1,L^Z_1%
\}$ then forms a representation of the algebra $\mathfrak{su}(2)$; i.e. they
are the logical qubit operators.

There is another way to see the action of these operators on the ground
states of $H_{\mathrm{cp}}$. Note that $H_{\mathrm{cp}}$ is time reversal
symmetric and the number of spin$-1/2$ particles in the system is $n^3$
(odd). Hence by Kramer's Theorem, each eigenspace has degeneracy which is a
multiple of $2$. The eigenstates come in pairs $(\ket{\lambda},\mho%
\ket{\lambda})$ where the anti-linear time reversal operator acts as $\mho%
\ket{\lambda}=K\mathcal{C}\ket{\lambda}$ where $\mathcal{C}$ is the complex
conjugation operation and $K=\prod_{i,j,k=1}^n(-i\sigma^y_{i,j,k})$. Because
the Hamiltonian is real, the eigenstates can be chosen real such that any
pair are given by $(\ket{\lambda},K\ket{\lambda})$. Now $K=-%
\prod_{i,j,k=1}^n
\sigma^z_{i,j,k}\prod_{i,j,k=1}^n\sigma^x_{i,j,k}=-[%
\prod_{k=1}^{(n-1)/2}V^Z_{2k}]L^Z_1 [\prod_{i=1}^{(n-1)/2}V^X_{2i}]L^X_1$.
But in a given stabilizer eigenspace $\mathcal{H}_{v^X,v^Z}$, the action of
this operation is $K=-(\prod_{i,k=1}^{(n-1)/2}v^X_{2i} v^Z_{2k})L^Z_1L^X_1$.
Restricting to the subspace $v^X_i=v^Z_k=1\forall i,k$, we have $K=-iL^Y_1$.
This then defines $(-i)$ times the logical $Y$ operation on that subspace.
From the commutation relations the operators $L^X_1$ and $L^Z_1$ are the
logical $X$ and $Z$ operations respectively.

A logical $\mathrm{CNOT}$ operation can be done transversally between two
code blocks. This follows by considering the action of the $\mathrm{CNOT}%
_{i,j}$ operation on the Pauli operators:
\begin{align}
\mathrm{CNOT}_{i,j}X_{i}\mathrm{CNOT}_{i,j} & = X_{i}X_{j}, & \mathrm{
CNOT}_{i,j}Z_{i}\mathrm{CNOT}_{i,j}&=Z_{i},\nonumber \\
\mathrm{CNOT}_{i,j}X_{j}\mathrm{CNOT}_{i,j} & = X_{j}, & \mathrm{CNOT}
_{i,j}Z_{j}\mathrm{CNOT}_{i,j}&=Z_{i}Z_{j},
\label{CNOT}
\end{align}%
Hence, by the group homomorphism, the joint stabilizer groups for the
control $I$ and target $J$ logical qubits $\mathcal{V}\times \mathcal{V}$ is
preserved under conjugation by $\mathrm{CNOT}^{\otimes n^{3}}=\mathrm{CNOT}%
_{I,J}$. This is easily checked by noting that even numbers of planar $%
L^{X},L^{Z}$ operators get mapped to even numbers of planar operators of the
same type. However the logical operators are acted upon nontrivially.
Specifically, we have the same relations as in Eq. \ref{CNOT} but with
logical operators replacing the physical qubit operators. Hence the
transversal $\mathrm{CNOT}$ is a logical $\mathrm{CNOT}$ on the code blocks.
Given the ability to generate arbitrary rotations $e^{i\phi L^{X}},e^{i\phi
L^{Z}}$ and the CNOT gate, and measurements of $L^{Z}$, exactly universal
quantum computation is allowed. 

The corresponding code is a $[[n^3,1,n]]$ code, i.e. it encodes $1$ logical
qubit in $n^3$ physical qubits with a distance $d=n$. This code can detect
up to $d-1$ arbitrary errors and correct $\lceil (n-1)/2\rceil$ errors which
is the maximal length of a an arbitrary error string with an unambiguous
action. Error correction is done by finding the minimum Hamming weight $n$
bit string consistent with the stabilizer measurements and applying single
spin(phase) flips on those planes corresponding $\sigma^{x(z)}$ error
locations.  An example
of a worst case error which saturates these numbers is the length $\ell$
error string $E=\prod_{i=i_0}^{i_0+\ell}\sigma^z_{i,j,k}$. The error string
has two boundaries which flips the sign of the stabilizers measurements $%
V^X_{i_0}$ and $V^X_{i_0+\ell}$. For $\ell<\lceil (n-1)/2\rceil$, the error
string creates two boundaries which are detected and appropriately
corrected. But for longer strings the error correction proceedure implements
a logical error on the code.

Some requirements for robust information processing in the above code are:

\begin{itemize}
\item Preparation of the system in the ground subspace of $H$. Presumably
can be accomplished by cooling the system to a pure separable state of a
local Hamiltonian $H_0=\sum_{i,j,k=1}^n\sigma^x_{i,j,k}$ then adiabatically
turning on $H$. It would be necessary to check that the adiabaticity
requirement was satisfied by estimating the gap of the time dependent
Hamiltonian $H(s)=(1-s)H_0+sH, s=t/T\in[0,1]$.

\item Projection onto a fiducial logical state in the ground subspace. This
can be done by measuring $m_Z=\langle L^Z_1\rangle$ and assigning the
logical state $\ket{0_L}(\ket{1_L})$ to outcome $m_Z=\pm 1$. Such a process
also allows measurement of the state in the logical $Z$ basis.

\item Encoding a quantum state and performing single qubit rotations. In
order to do so it is necessary to be able to implement single logical qubit
operations. To generate the continuous group SU$(2)$, the continuous gate
set $\{e^{i\xi L^X_1},e^{i\zeta L^Z_1}\}$ suffices. One may rather demand
only a discrete gate set that generates a group dense in SU$(2)$. One such
gate library that can be done fault tolerantly is $\{L^h=e^{i\frac{\pi}{2}%
L^Z_1}e^{i\frac{\pi}{4}L^Y_1},e^{-i\frac{\pi}{8} L^Z_1},\mathrm{CNOT}\}$.
Since the $L^Y$ is a product of $\sigma^y$ operators on all qubits, single
logical qubit gates could be performing by emersing the crystal in one
direction into the cavity, first one plane to generate gates via $L^Z_1$ and
then the entire crystal to generated gates from $L^Y$.
The $\mathrm{CNOT}$ can be performed transversally between two code blocks
by performing physical $\mathrm{CNOT}$ gates in parallel with $n^3$ control
physical qubits in one code block acting on $n^3$ target physical qubits in
the target code block. This one step parallel operation is difficult to do
with local operations, however one could perform the gate locally by
performing parallel $\mathrm{CNOT}$ gates between the $n^2$ physical qubits
in the bottom most $\hat{x}-\hat{y}$ plane of the control logical qubit and
the $n^2$ target physical qubits in the top most $\hat{x}-\hat{y}$ plane of
the target logical qubit. A series of $n-1$ such steps where the $\hat{x}-%
\hat{z}$ planes of the logical qubits are cyclically shifted realizes the logical $\mathrm{CNOT}$. Each
cyclic shifts can be done in a linear number of parallel planar $\mathrm{SWAP%
}$ gates.


\item Measuring stabilizer operators is necessary to detect and ultimately
correct errors. This demands measuring the set of $2(n-1)$ independent
stabilizer generators $\{V_{i}^{X},V_{k}^{Z}\}$ which are nearest neighbor
planes of products of all $\sigma ^{x}$ or all $\sigma ^{z}$ operations.
\end{itemize}

\section{Alternate derivation of dynamics during the geometric phase gate}

\label{AppA}

The evolution of the joint system of spins and field in a coherent state
basis was derived in Sec. \ref{gpfid} by integrating the equation of motion
including cavity decay. Here we provide an alternative derivation using
characteristic equation for the state. We begin by transforming the
evolution in Eq. \ref{rhodot} to an interaction picture via
\[
\rho_I (t)=e^{iV_Z t}\rho(t)e^{-iV_Z t},\quad a_I(t)=e^{iV_Z t}a(t)e^{-iV_Z
t},
\]
we have
\[
\dot{\rho}_I(t)=\kappa (a_I \rho_I a_I^{\dagger}-\frac{1}{2}%
a_I^{\dagger}a_I\rho_I-\frac{1}{2}\rho_I a_I^{\dagger}a_I).
\]
Now $a_I(t)=e^{-i2g_Z J^z t}a$ hence,
\[
\dot{\rho}_I(t)=\kappa (a e^{-i2g_Z J^z t}\rho_I e^{i2g_Z J^z t} a^{\dagger}-%
\frac{1}{2}a^{\dagger}a\rho_I-\frac{1}{2}\rho_I a^{\dagger}a).
\]
In the interaction picture:
\[
A^{M_J,M^{\prime }_J}_I(t)=\ket{\Lambda,J,M_J}\bra{\Lambda',J',M'_J}%
(t)\otimes \ket{\tilde{\alpha}_{M_J}(t) }\bra{\tilde{\beta}_{M'_J}(t)},
\]
where: $\tilde{\alpha}_{M_J}(t)=\alpha_{M_J}(t)e^{i2g_Z t M_J},\tilde{\beta}%
_{M^{\prime }_J}(t)=\beta_{M^{\prime }_J}(t)e^{i2g_Z t M^{\prime }_J})$, To
derive the evolution during decay we use the characteristic function
\[
X(t)=\mbox{Tr}_F [A^{M_J,M^{\prime }_J}_I(t)e^{\lambda
a^{\dagger}}e^{-\lambda^{\ast}a}],
\]
(where the trace is taken over the field) such that
\begin{align*}
\dot{X}(t) & = \mbox{Tr}_F[\dot{A}^{M_J,M^{\prime }_J}_I(t) e^{\lambda
a^{\dagger}}e^{-\lambda^{\ast}a}] \\
& = \kappa \mbox{Tr}_F [(e^{-i2g_Z t (M_J-M^{\prime }_J)}
aA^{M_J,M^{\prime }_J}_I(t)a^{\dagger} \\
& \phantom{=} -\frac{1}{2}a^{\dagger}a A^{M_J,M^{\prime }_J}_I(t) -\frac{1}{2}
A^{M_J,M^{\prime }_J}_I(t)a^{\dagger}a) e^{\lambda
a^{\dagger}}e^{-\lambda^{\ast} a}].%
\end{align*}
Using the relations:
\[
e^{-\lambda^{\ast}a}a^{\dagger}=(a^{\dagger}-\lambda^{\ast})e^{-\lambda^{%
\ast}a},\quad ae^{\lambda a^{\dagger}}=e^{\lambda a^{\dagger}}(a+\lambda),
\]
we obtain
\begin{align}
\dot{X} & = -\kappa (e^{-i2g_Z t (M_J-M^{\prime }_J)}-1)\frac{\partial^2}{%
\partial \lambda^{\ast}\partial\lambda}X-\frac{\kappa}{2}\Big(\lambda^{\ast}%
\frac{\partial X}{\partial \lambda^{\ast}}+\lambda \frac{\partial X}{%
\partial \lambda}\Big) \nonumber\\
& = \kappa (e^{-i2g_Z t (M_J-M^{\prime }_J)}-1)\tilde{\alpha}_{M_J}(t)%
\tilde{\beta}_{M^{\prime }_J}^{\ast}(t) X \nonumber\\
& \phantom{=}  -\frac{\kappa}{2}\Big(\tilde{\beta}_{M^{\prime }_J}^{\ast}(t)\lambda-%
\tilde{\alpha}_{M_J}(t)\lambda^{\ast}\Big)X.%
\label{Xd}
\end{align}

The equation of motion for the operator $X$ can be solved by the method of
characteristics. For the states of interest here we make the ansatz:
\begin{equation}
X(t)=C(t)e^{-\lambda ^{\ast }\tilde{\alpha}_{M_{J}}(t)}e^{\lambda \tilde{%
\beta}_{M_{J}^{\prime }}^{\ast }(t)}\ket{\Lambda,J,M_J}\bra{\Lambda',J',M'_J}%
.  \label{anz}
\end{equation}%
Evaluating the time derivative and setting this equal to Eq. \ref{Xd} we
find
\begin{align*}
\tilde{\alpha}_{M_{J}}(t) & = e^{-\kappa t/2}\alpha _{M_{J}},\\
\tilde{\beta}_{M_{J}^{\prime }}(t) &=e^{-\kappa t/2}\beta _{M_{J}^{\prime }}, \\
X(t) & = \langle \beta _{M_{J}^{\prime }}|\alpha _{M_{J}}\rangle
^{(1-e^{-\kappa t})}e^{c(t)}e^{-\lambda ^{\ast }\alpha e^{-\kappa
t/2}}e^{\lambda \beta ^{\ast }e^{-\kappa t/2}} \\
&  \quad\times\langle \beta _{M_{J}^{\prime }}e^{-\kappa t/2}|\alpha
_{M_{J}}e^{-\kappa t/2}\rangle \ket{\Lambda,J,M_J}\bra{\Lambda',J',M'_J}.%
\end{align*}%
Hence,
\begin{align*}
c(t) & = \int_{0}^{t}dt^{\prime }\kappa \lbrack \tilde{\alpha}%
_{M_{J}}(t^{\prime })\tilde{\beta}_{M_{J}^{\prime }}^{\ast }(t^{\prime
-i2g_{Z}t^{\prime }(M_{J}-M_{J}^{\prime })}-1)] \\
& = \frac{1}{\kappa+i2g_{Z}(M_{J}-M_{J}^{\prime })}\left\{\alpha _{M_{J}}\beta _{M_{J}^{\prime }}^{\ast }[(e^{-\kappa
t}-1)i2g_{Z}(M_{J}-M_{J}^{\prime })\right.\\
&\quad +\left.\kappa e^{-\kappa
t}(1-e^{-i2g_{Z}t(M_{J}-M_{J}^{\prime })})]\right\},%
\end{align*}
where we have chosen the integration constant so that $c(0)=0$. Notice that
for $g_{Z}=0$, i.e. pure decay, then $c(t)=0$. This should be the case as
then $\ket{\alpha_{M_J}}\bra{\beta_{M'_J}}\rightarrow \langle \beta
_{M_{J}^{\prime }}|\alpha _{M_{J}}\rangle ^{(1-e^{-\kappa t})}%
\ket{\alpha_{M_J} e^{-\kappa
t/2}}\bra{\beta_{M'_J} e^{-\kappa t/2}}$. Furthermore, for $%
M_{J}=M_{J}^{\prime }$ then $c(t)=0$. To account for all phases and decay we
introduce:
\[
d(t)=c(t)-\frac{1-e^{-\kappa t}}{2}(|\alpha _{M_{J}}|^{2}+|\beta
_{M_{J}^{\prime }}|^{2}-2\alpha _{M_{J}}\beta _{M_{J}^{\prime }}^{\ast }).
\]%
Finally we arrive at
\begin{align*}
\rho (t) & = e^{\mathcal{L}t}\rho (0) \\
& = \displaystyle{\sum_{\Lambda ,J,M_{J},\Lambda ^{\prime },J^{\prime
},M_{J}^{\prime }}}\sigma _{\Lambda ,J,M_{J},\Lambda ^{\prime },J^{\prime
},M_{J}^{\prime }}e^{d(t)}\ket{\Lambda,J,M_J}\bra{\Lambda',J',M'_J} \\
& \quad \otimes \ket{\alpha e^{-\kappa t/2}e^{-i2g_ZtM_J}}\bra{\beta e^{-\kappa
t/2}e^{-i2g_Zt M'_J }},
\end{align*}
which is the same evolution as derived in Eq. \ref{evolved}.

\end{document}